\documentclass[prd,floatfix,showpacs,preprintnumbers,nofootinbib,superscriptaddress]{revtex4}
\usepackage{amsmath,amssymb}
\usepackage[usenames,dvipsnames]{color}
\usepackage{graphicx}
\usepackage{subfigure}

\newcommand\bef{\begin{figure}}
\newcommand\eef[1]{\label{fg:#1}\end{figure}}
\newcommand\besf{\begin{subfigure}}
\newcommand\eesf[1]{\label{sfg:#1}\end{subfigure}}
\newcommand\beq{\begin{equation}}
\newcommand\eeq[1]{\label{#1}\end{equation}}
\newcommand\beqa{\begin{eqnarray}}
\newcommand\eeqa[1]{\label{#1}\end{eqnarray}}
\newcommand\bet{\begin{table}}
\newcommand\eet[1]{\label{tb:#1}\end{table}}
\newcommand\best{\begin{subtable}}
\newcommand\eest[1]{\label{stb:#1}\end{subtable}}
\newcommand\betb{\begin{center}\begin{tabular}}
\newcommand\eetb{\end{tabular}\end{center}}
\newcommand\beit{\begin{itemize}}
\newcommand\eeit{\end{itemize}}

\newcommand\fgn[1]{Figure \ref{fg:#1}}
\newcommand\eqn[1]{eq.\ (\ref{#1})}

\newcommand\tbn[1]{Table \ref{tb:#1}}

\newcommand\cdof{\chi^2/\mathrm{DOF}}

\begin{document}
\title{Spectroscopy of triply-charmed baryons from lattice QCD}
\author{M.\ \surname{Padmanath}}
\email{padmanath@theory.tifr.res.in}
\affiliation{Department of Theoretical Physics, Tata Institute of Fundamental
         Research,\\ Homi Bhabha Road, Mumbai 400005, India.}

\author{Robert\ G.\ \surname{Edwards}}
\email{edwards@jlab.org}
\affiliation{Jefferson Laboratory, 12000 Jefferson Avenue,  Newport News, VA 23606, USA}

\author{Nilmani\ \surname{Mathur}}
\email{nilmani@theory.tifr.res.in}
\affiliation{Department of Theoretical Physics, Tata Institute of Fundamental
         Research,\\ Homi Bhabha Road, Mumbai 400005, India.}

\author{Michael\ \surname{Peardon}}
\email{mjp@maths.tcd.ie}
\affiliation{School of Mathematics, Trinity College, Dublin 2, Ireland}

\collaboration{For the Hadron Spectrum Collaboration}

\begin{abstract}
The spectrum of excitations of triply-charmed baryons is computed
using lattice QCD including dynamical light quark fields. Calculations
are performed on anisotropic lattices with temporal and spatial
spacings $a_{t} = 0.0351(2)$ fm and $a_s \sim 0.12$ fm respectively and
with pion mass of about 390 MeV. The spectrum obtained has baryonic
states with well-defined total spin up to $\frac{7}{2}$ and the
low-lying states closely resemble the expectation from models with an
$SU(6)\times O(3)$ symmetry. Energy splittings between extracted
states, including those due to spin-orbit coupling in the heavy quark
limit are computed and compared against data at other quark masses.
\end{abstract}

\maketitle

\section{Introduction}

The charmonium spectrum has been studied in detail both in experiments and 
theoretical calculations over many years. This has provided crucial input into 
our understanding of the nature of the strong interaction. In contrast, 
while charm baryons can provide similar insight, they have not yet been studied
so thoroughly. Only a handful of charmed baryons have been discovered and a 
reliable determination of the quantum number of most of the observed states
has not been made~\cite{PDG}.  Only very recently
have a few excited singly charmed baryons been observed and the discovery 
status of the doubly charmed baryons remains unsettled. While the SELEX 
experiment observed doubly charmed $\Xi_{cc}(ccu)$
baryons~\cite{Mattson:2002vu,Russ:2002bw, Ocherashvili:2004hi}, these
have not been confirmed either by BABAR~\cite{Aubert:2006qw} or
Belle~\cite{Chistov:2006zj}. Along with the well-established triply flavoured  
$\Delta (uuu)$ and strange $\Omega(sss)$ baryons, QCD predicts similar
states built from charm quarks, the triply-charmed baryon, $\Omega_{ccc}$. 
Such a state has yet to be observed. Beside $\Omega_{ccc}$, QCD also
predicts many other triply-charmed baryons, which can be considered as the 
baryon analogues of charmonia. While it has been pointed out that 
semileptonic decay processes such as $\Omega_{ccc}^{++} \rightarrow
\Omega_{sss}^{-} + 3\mu^{+} + 3\nu_{\mu}$ and $\Omega_{ccc}^{++} \rightarrow
\Omega_{sss}^{-} + 3\pi^{+}$ possibly can offer 
signature for a $``ccc"$ event~\cite{Bjorken:1985ei}, the production of
triply-charmed baryons in current charm factories is difficult
(see Refs.~\cite{Chen:2011mb, GomshiNobary:2005ur} for triply-charmed baryon 
production).  However, it is expected that the large statistical sample 
collected at the LHCb experiment, the PANDA experiment at the FAIR facility, 
Belle II at KEK and BES III may be able to provide some information on 
triply-charmed baryons, along with other baryons containing charm quarks.

The triply-charmed baryons may provide a new window for understanding the
structure of baryons, as pointed out by Bjorken several
years ago~\cite{Bjorken:1985ei}.  A comprehensive study of the excitation 
spectra of these
states, where the complications of light-quark interaction are absent, can 
provide information about the quark confinement mechanism as well as 
elucidating our knowledge about the nature of the strong force by providing a 
clean probe of the interplay between perturbative and nonperturbative 
QCD~\cite{Brambilla:2010cs}. On the theoretical side, one expects that 
potential models will be able to describe triply-charmed baryons to a similar 
level of precision as their success in charmonia. Just as the quark-antiquark 
interactions are examined in charmonia, these studies will probe the 
quark-quark interactions in the charm sector. The spectra for triply-charmed 
baryons have been studied theoretically by 
non-relativistic~\cite{Roberts:2007ni, Vijande:2004at} and 
relativistic~\cite{Martynenko:2007je, Migura:2006ep} quark models, quark 
models in the Faddeev formalism~\cite{SilvestreBrac:1996bg}, the bag 
model~\cite{Hasenfratz:1980ka}, effective field theory with potential 
NRQCD~\cite{Brambilla:2004jw, Brambilla:2005yk, Brambilla:2009cd, 
  LlanesEstrada:2011kc},  
 heavy quark spin symmetry~\cite{Flynn:2011gf}, variational 
 method~\cite{Jia:2006gw}, QCD sum rules~\cite{Zhang:2009re, Wang:2011ae}, 
 Regge phenomenology~\cite{Guo:2008he} and the one-gluon-exchange 
 model~\cite{Korner:1994nh}. 
However, in the absence of any experimental discovery the only way to test 
these model-dependent as well as perturbative approaches is to compare these 
with the results from non-perturbative lattice QCD calculations.  By using 
lattice QCD, several calculations have already been performed to compute the 
ground state triply-charmed baryon, $\Omega_{ccc}(J^P = \frac32^{+}) $, 
including quenched~\cite{Chiu:2005zc} as well as full 
QCD~\cite{Alexandrou:2012xk, Briceno:2012wt, Basak:2012py, Durr:2012dw, Namekawa:2013vu}. 

A quantitative description of the spectra of triply-charmed baryons 
using a non-perturbative method such as lattice QCD is thus important for a 
number of reasons. Firstly, as mentioned above, it will be interesting to 
compare the spectra of triply-charmed baryons computed from a first principles 
method to those obtained from potential models which have been very successful 
for charmonia. Secondly, all results from such first principles calculations 
will be predictions and thus naturally can provide crucial inputs to future 
experimental discovery.  Given this significance of triply-charmed baryons, it 
is desirable to study these states comprehensively using lattice QCD. All 
previous lattice calculations have studied only the 
$\Omega_{ccc}(J^P = \frac32^{+})$ ground state. However, it is expected that 
more information about interactions between multiple charm quarks can be 
obtained by computing the excited state spectra of 
$\Omega_{ccc}(J^P = \frac32^{+})$  baryons, including in particular the 
spin-dependent energy splittings, as well as by studying similar spectra for 
other triply-charmed baryons with all spin and parity combinations.  In this 
work, we make 
the first attempt towards this goal and compute the excited state spectra of 
triply-charmed baryons using dynamical lattice QCD. The ground states for each 
spin-parity channel up to spin 7/2 are computed along with a number of their
excited states and several spin-dependent energy splittings. 
Similar studies of singly and doubly-charmed baryons will be reported in 
subsequent 
publications.

We use a well defined procedure developed and employed extensively for 
extracting excited states for light
mesons~\cite{Dudek:2010wm, Dudek:2009qf, Dudek:2010ew}, 
mesons containing charm quarks~\cite{Liu:2012ze, Moir:2013ub}, and light and 
strange baryons~\cite{Edwards:2011jj, Dudek:2012ag, Edwards:2012fx}.  This
method uses anisotropic lattice configurations 
with the light quark dynamics 
included~\cite{Edwards:2008ja,Lin:2008pr}.
The anisotropic discretisation helps us to obtain better resolution of the 
correlation functions, which is very helpful for the extraction of excited 
states.
Furthermore, we construct a large set of baryonic operators in the 
continuum and then subduce them to various lattice irreducible
representations to obtain lattice
operators~\cite{Edwards:2011jj}. These operators transform as
irreducible representations (irreps) of $SU(3)_F$ symmetry for
flavour, $SU(4)$ symmetry for Dirac spins of quarks and $O(3)$
symmetry for orbital angular momenta. Finally, using a novel
technique called ``distillation''~\cite{Peardon:2009gh}, correlation functions 
of these operators are computed and the variational method is utilized to
extract excited energies as well as to reliably determine the spins of
these states.

The layout of the remainder of the paper is as follows.  In the next
section, we describe the details of our numerical methods mentioning
first the details of lattices used and then the construction of the
lattice interpolating operators in subsection IIB. In
subsection IIC, we detail the generation and analysis of
baryon correlation functions by distillation and variational methods
and then in IID we discuss our procedure of identifying
the continuum spins of the extracted states; subsection
IIE discusses about the lattice artefacts related to this work. In section 
III we present our results and give details of energy
splittings in subsection IIIA. Finally, a summary of the work is
presented in section IV.

\section{Lattice methodology}

In this section, the details of our Monte Carlo calculation of triply charmed
baryon excitations using lattice QCD are described.  In recent years the 
Hadron Spectrum Collaboration has exploited a dynamical anisotropic lattice 
formulation to extract highly excited hadron spectra. In this approach a 
lattice with a much finer temporal spacing than in the spatial directions is 
employed. The high temporal resolution means many time slices are available to 
compute the signal for highly excited states which decay very rapidly at large 
Euclidean time separations leading to a increase in the noise-to-signal ratio. 
The anisotropic lattice achieves this resolution while avoiding the 
computational cost which would come from reducing the spacing in all directions. 

\subsection{The lattice action}

The tree-level 
Symanzik-improved gauge action and the anisotropic Shekholeslami-Wohlert 
fermion action with tree-level tadpole improvement and three-dimensional 
stout-link smearing of gauge fields are used. More details of the formulation 
of actions as well as the techniques used to determine the anisotropy 
parameters can be found in Refs.~\cite{Edwards:2008ja, Lin:2008pr}.
\bet[h]
\centering
\betb{cccc|ccc}
\hline
Lattice size & $a_t m_\ell$   & $a_t m_s$ & $N_{\mathrm{cfgs}}$ & $m_\pi/$MeV & $m_K/m_{\pi}$ &$a_t m_\Omega$ \\\hline
$16^3 \times 128$ & $-0.0840$ & $-0.0743$ & 96 & 391&1.39        & $0.2951(22)$      
\\\hline
\eetb
\caption{Properties of the gauge-field ensembles used. $N_{\mathrm{cfgs}}$  is the number of gauge-field 
configurations. }
\eet{lattices}

The lattice action parameters of the gauge-field ensembles used 
this work are given in \tbn{lattices}. As mentioned in
Ref.~\cite{Edwards:2011jj}, the temporal lattice spacing $a_t$ was
determined by equating the $\Omega$-baryon mass measured on these
ensembles $a_tm_{\Omega} = 0.2951(22)$ with the physical value, 
$m_{\Omega} = 1672.45(29)$ MeV. This leads to $a_t^{-1} = 5.67(4)$ GeV and
with an anisotropy of close to 3.5, $a_s =$ 0.12 fm. This gives a spatial 
extent of about $1.9$ fm, which should be sufficiently large for a study of 
triply-charmed baryons. 

The details of the charm quark action used for
this study are given in Ref.~\cite{Liu:2012ze}. The action parameters for 
the charm quark are obtained by ensuring the mass of the $\eta_c$ meson 
takes its physical value and its dispersion relation at low momentum is 
relativistic. As mentioned in~\cite{Liu:2012ze}, it is expected that the 
effects due to the absence of dynamical charm quark fields in this calculation
will be small. 

\subsection{Baryon operator construction}

We follow the same methods described in Ref.~\cite{Edwards:2011jj}
for the construction of baryon interpolating operators on the
lattice. Baryons are colour singlets made of three quarks and so have
anti-symmetric wave functions in colour space. Since they are fermions,
their interpolating operators excluding the colour part should be 
totally symmetric combinations of all the quark labels representing
flavour, spin and the spatial structure. Our construction of baryon 
interpolating 
operators proceeds in two stages; first, a set of continuum operators with 
well-defined continuum spin is defined. These operators are then subduced 
into irreducible representations of the double-cover octahedral group.

The overall spin and parity of a continuum baryon interpolating operator 
can be found by decomposing it into a combination of three components;
\beq O^{[J^P]} = \left[
  \mathcal{F}_{\Sigma_F}\otimes\mathcal{S}_{\Sigma_S}\otimes\mathcal{D}_{\Sigma_
D}
  \right] ^{J^P}, 
\eeq{genericbaryon} 
where $\cal{F}$, $\cal{S}$ and $\cal{D}$ represent flavour, Dirac spin and 
spatial structure respectively while the subscripts $\Sigma_i$ specify the 
permutation symmetry in the respective subspaces. The details of these 
permutation symmetries and their combinations are given in
Refs.~\cite{Edwards:2011jj, Edwards:2012fx}. After combining them
together, we get interpolating fields with specific spin and
parity which are overall symmetric under permutations.
The flavour combination $\mathcal{F}_{\Sigma_F}$ of triply-charmed
operators must be totally symmetric and is the same as that of the $I_Z = \pm 
3/2$ part of the $\Delta$ and the $\Omega$ baryon operators. Consequently the remaining 
spin and spatial part must be combined together in a symmetric combination to 
form an overall symmetric interpolating operator.
The spin symmetry combinations $\mathcal{S}_{\Sigma_S}$ in Dirac space
can be obtained by combining the $\rho$ and $\sigma$ spins for a quark
field $\psi^{\mu} = q^{\rho}_{\sigma}$, as explained in
Ref.~\cite{Edwards:2011jj}.  
Only creation operators formed from the upper two components of the 
four-component Dirac spinor appear at leading order in a velocity expansion. 
Those based on the lower components of Dirac spinors are relativistic.

The spatial symmetry combinations $\mathcal{D}_{\Sigma_D}$ depend on
the gauge covariant derivatives acting on the three quark fields. As
in our previous studies~\cite{Edwards:2011jj, Edwards:2012fx}, 
we include up to two covariant derivatives.
These are combined to transform irreducibly with orbital angular momentum, 
$L$, giving access to $L=0,1$ and $2$.
Without derivatives, only the totally symmetric Dirac spin structure is 
allowed. The single derivative structures can have mixed symmetry or mixed
antisymmetry and hence they can only be combined with Dirac spin structures 
with mixed symmetry and mixed antisymmetry respectively. With up to one 
derivative it is possible to construct operators with spins up to $\frac32$ 
while for baryons with flavour-mixed symmetry it is up to $\frac52$.  
However, two derivative projection
operators can be combined with all spin symmetry combinations
$\mathcal{S}_{\Sigma_S}$ in Dirac space which enables states
with spins up to $\frac72$.  In \tbn{n_operators}, we show allowed
spin-parity patterns of operators using the upper quark spinors combined with
up to two covariant derivatives.  Note that with the non-relativistic Dirac 
spin components alone, it is not possible to construct a negative parity state 
beyond spin $\frac32^-$ even with operators that include two derivatives. This 
limited the extraction of spin $\frac52^-$ and spin $\frac72^-$ states in
Ref.~\cite{Meinel:2012qz}. Use of relativistic operators along with
non-relativistic ones enable us to extract these negative parity
states along with more excited states.

\bet[h]
\centering
\betb{ccc|cccc|cc}
    \hline
      &     &     & \multicolumn{6}{c}{$N(J^P)$}\\
    \hline
$N_D$ & $L$ & $S$ & $\frac12^+$ & $\frac32^+$ & $\frac52^+$ & $\frac72^+$ & 
                    $\frac12^-$ & $\frac32^-$ \\[1mm]
\hline 
\hline 
0 & 0 & $\frac32$ &   & 1 &   &   &   &   \\[1mm]
\hline 
1 & 1 & $\frac12$ &   &   &   &   & 1 & 1 \\[1mm]
\hline 
2 & 0 & $\frac12$ & 1 &   &   &   &   &   \\[1mm]
2 & 0 & $\frac32$ &   & 1 &   &   &   &   \\[1mm]
{\bf 2} & {\bf 1} & $\mathbf{\frac12}$ & {\bf 1} & {\bf 1} &   &   &   &   \\[1mm]
2 & 2 & $\frac12$ &   & 1 & 1 &   &   &   \\[1mm]
2 & 2 & $\frac32$ & 1 & 1 & 1 & 1 &   &   \\[1mm]
\hline 
\eetb
\caption{The number of operators of a given $J^P$ that can be constructed from
up to two derivatives acting on non-relativistic quark spinors. $N_D$ indicates
  the number of covariant derivatives, $S$ indicates the total spin of the 
  quarks and $L$ indicates the total orbital angular momentum. The row indicated
  in bold face contains the two-derivative ``hybrid'' operators, which 
  vanish in the absence of a gluon field. }
\eet{n_operators} 
With two derivatives, a subset of operators with $L = 1$ in mixed-symmetric
and mixed-antisymmetric combinations \cite{Dudek:2012ag} are identified as 
hybrid operators because they vanish in the absence of a gluon field. 
Note that a colour-singlet object $[(qqq)_{8_c}G_{8_c}]_{1_c}$ can be 
constructed through a combination of three quarks in a colour octet
with a gluon field, G. \tbn{n_operators} shows the pattern of
states expected to be created by two-derivatives operators
built from non-relativistic quark spinors. As will be observed later, 
operators which incorporate the gluon field-strength tensor are essential in 
obtaining some states in the spectrum. 

To define a creation operator that respects the symmetries of the lattice, 
continuum operators with definite spin and parity are subduced into the irreps 
of the cubic group. The three irreps of the double-valued representations of the
octahedral group for half-integer spins are $G_1$, $G_2$ and $H$. The
details of this subduction procedure was discussed in
Ref.~\cite{Edwards:2011jj}. After completing the subduction procedure, 
the number of operators used for this study are shown in
\tbn{operators}. In each irrep, the same number of operators for
both positive and negative parities, denoted by the subscript $g$ and
$u$, respectively are used. This table also gives the number of
operators made exclusively from non-relativistic quark spinors (NR) and those 
hybrid operators which vanish in the absence of a gluon field. 

We have followed the same naming convention as Ref.~\cite{Edwards:2011jj}. Since there are only charm valence quarks in this calculation we do not refer 
to flavour and name an operator according to its spin and spatial structure. 
For example, in the 
operator $O = [(\frac32^+_{1,S}) \otimes D^{[2]}_{L=2,S}]^{\frac72} H_{g1}$, 
the label $(\frac32^+_{1,S})$ indicates the quark spins are combined to form 
the first (1) embedding of the $H_g$ (3/2) irrep in the symmetric (S) 
combination, $D^{[2]}_{L=2,S}$ represents two covariant derivatives with two 
units of angular momentum in the symmetric combination, the $^{\frac72}$ part 
of the name indicates the continuum spin $J=7/2$ and $H_{g1}$ indicates the 
lattice operator transforms as the first row of the subduced irrep $H_g$.
\bet[h]	
\centering
\betb{ c | c  c  c  c  c  c }
\hline 
       &\multicolumn{2}{c}{$G_1$}&\multicolumn{2}{c}{H}&\multicolumn{2}{c}{$G_2$
} \\ \cline{2-7}
       &   $g$    &     $u$      &    $g$   &    $u$   &     $g$   &    $u$     
  \\ \hline
Total  &    20    &      20      &     33   &     33   &      12   &     12     
  \\
Hybrid &     4    &       4      &      5   &      5   &       1   &      1     
  \\
NR     &     4    &       1      &      8   &      1   &       3   &      0     
  \\
\hline
\eetb
\caption{
The number of lattice operators obtained after subduction to various 
irreps of operators with up to two covariant derivatives. The number of 
non-relativistic (NR) and hybrid operators for each irreps and for both 
parities are given.}
\eet{operators}

\subsection{Variational analysis of correlation functions}

We employ a variational method to extract the spectrum of baryon
states from the matrix of correlation functions calculated by using
the large basis of interpolating operators described
above. For operators $i$ and $j$, in a given irrep $\Lambda$, we
calculate the matrix of correlation functions 
\beq
C^{\Lambda}_{ij}(t_f-t_i) = \langle 0
|\bar{O}_i(t_f)O_j(t_i)|0\rangle.  
\eeq{2pt} 
between a baryon creation operators at
time $t_i$ and an annihilation operator at time $t_f$.  An efficient way to
construct this correlation matrix for the large basis of interpolating
operators utilized in this work is through the {\it distillation method}, 
detailed in Ref.~\cite{Peardon:2009gh} and utilized in various calculations by
the collaboration. Distillation is a smearing method that defines a linear
smoothing operator with support only in a small space of physically relevant
vectors.  As with other recent calculations by the collaboration, we use the
low-lying eigenvectors of the gauge-covariant three-dimensional Laplacian in
this calculation.  The correlation function of Eq.~\ref{2pt} then factorises,
enabling efficient computation of all elements of the matrix. 
The distillation method is useful for these calculations since 
it provides a technique to determine the full matrix of correlation functions 
for any number of {\it smeared} operators at both source and sink. For this 
work, the distillation method was realized by constructing the smearing
operator from the lowest 64 eigenvectors of the Laplacian. Correlation 
functions are then constructed using a set of four time sources per gauge 
configuration.

The variational method enables a reliable extraction of the spectrum beyond 
the ground state. The method \cite{Michael:1985ne,Luscher:1990ck} proceeds by 
solving a generalized eigenvalue problem of the form
\beq C_{ij}(t)v_{j}^{n} =
\lambda_{n}(t,t_0)C_{ij}(t_0)v_{j}^{n},
\eeq{}
where $v^{n}$ is the $n$-th eigenvector 
and the eigenvalues, $\lambda_{n}(t,t_0)$ are termed the 
{\it principal correlators} and obey
\beq
\lim_{t-t_0\rightarrow\infty} \lambda_n(t,t_0) = e^{-E_n (t-t_0)}
\eeq{}
with $E_n$ the energy of the $n$-th excited state. An
appropriate reference time-slice $t_0$ is chosen for diagonalization
as described in Refs.~\cite{Dudek:2007wv,
  Dudek:2010wm,Dudek:2009qf}, which corresponds to a minimum of a $\chi^2$-like quantity as defined in Ref.~\cite{Dudek:2007wv}. We extract the energy of a state by
fitting the dependence of $\lambda_{n}$ on $t-t_0$
to the form, 
\beq 
   \lambda_n(t,t_0) = (1-A_n)e^{-m_n(t-t_0)} + A_ne^{-m'_n(t-t_0)},
\eeq{vf} 
with three fit parameters $m_n, m'_n$ and $A_n$. 
As with our previous studies, we find that allowing a second exponential 
stabilizes the fits and the resulting second exponential decreases rapidly 
with large $t_0$. All the excited states in our fits are found to be less than the second exponential $m'_n$. In \fgn{hgpc}, we plot
some examples of fits to the principal correlators in irrep $H_g$, where
the fitted states will be identified with $J^{P} = \frac32^+$.  The fits
approach the constant value, $1-A_n$, for large $t$, and they approach
one if a single exponential dominates. This is seen to be the case for most 
of our fits. In \tbn{principal_correlator} we show fit results for the lowest three states in each irrep. While fitting principal correlators we used 
a standard $\chi^2$ minimization procedure incorporating the measured data 
covariance.
\bef[tbh] \centering
\hspace*{-0.25in} 
\includegraphics[scale=1.35]{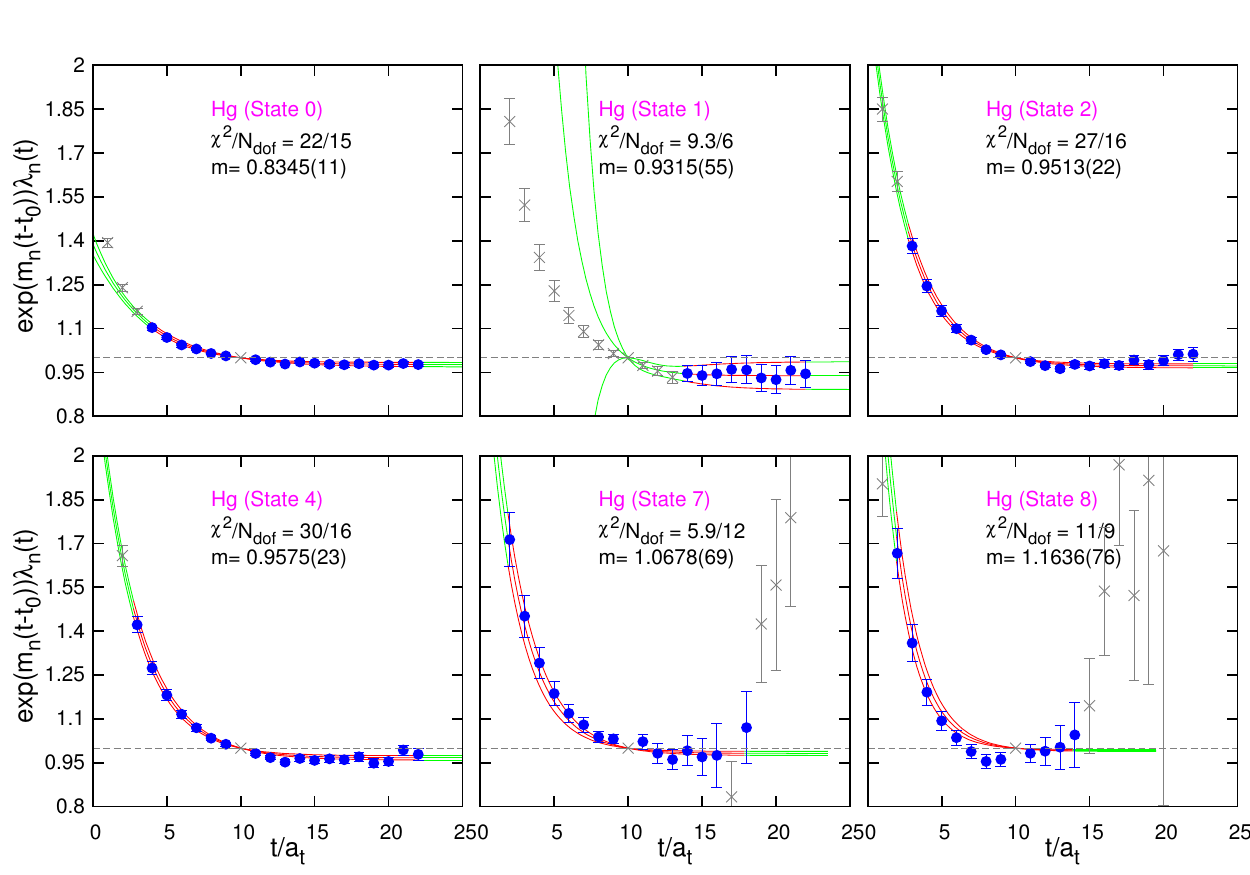}
\caption{Principal correlator fits for six states in irrep $H_{g}$
  that are identified as $J^P = 3/2^+$. Fits are obtained using
  \eqn{vf}.  Data points are obtained from
  $e^{m_n(t-t_0)}\lambda_n(t)$ and the lines show the fits and one
  sigma-deviation according to the fitting form
  $e^{m_n(t-t_0)}\lambda_n(t) = 1-A_n +A_n e^{-(m'_n-m_n)(t-t_0)}$,
  with $t_0 = 10$; the grey points are not included in the fits.}
\eef{hgpc}
\bet[h]
\centering
\betb{ c | c | c | c | c | c | c | c }
Irrep    & \multicolumn{3}{c}{positive parity}   \vline &&  \multicolumn{3}{c}{negative parity}   \\ \hline
         & $a_t~m_n$   &   Range & $\cdof$  && $a_t~m_n$   &   Range & $\cdof$ \\ \hline
$G_{1}$  & 0.9460 (23) &  [3-22] & 1.30     && 0.8970 (16) &  [4-22] & 1.44    \\
         & 0.9470 (24) &  [3-22] & 0.94     && 0.9832 (54) &  [4-22] & 0.77    \\
         & 0.9505 (23) &  [3-22] & 0.62     && 0.9866 (76) &  [4-22] & 2.54    \\ \hline
$H$      & 0.8345 (11) &  [4-22] & 1.49     && 0.8977 (23) &  [4-22] & 0.97    \\
         & 0.9315 (55) & [14-22] & 1.55     && 0.9558(259) &  [5-20] & 1.20    \\
         & 0.9513 (22) &  [3-22] & 1.62     && 0.9920 (53) &  [4-22] & 0.76    \\ \hline
$G_{2}$  & 0.9523 (18) &  [1-22] & 1.67     && 0.9189(217) &  [5-22] & 1.36    \\
         & 0.9524 (25) &  [3-22] & 1.38     && 0.9812(117) & [12-22] & 1.96    \\
         & 0.9529 (32) &  [4-22] & 1.56     && 0.9977 (55) &  [4-22] & 0.88
\eetb
\caption{Fit results from principal correlators of the lowest three states in all the irreps.}
\eet{principal_correlator}
\bef[b!]
\includegraphics[scale=0.3]{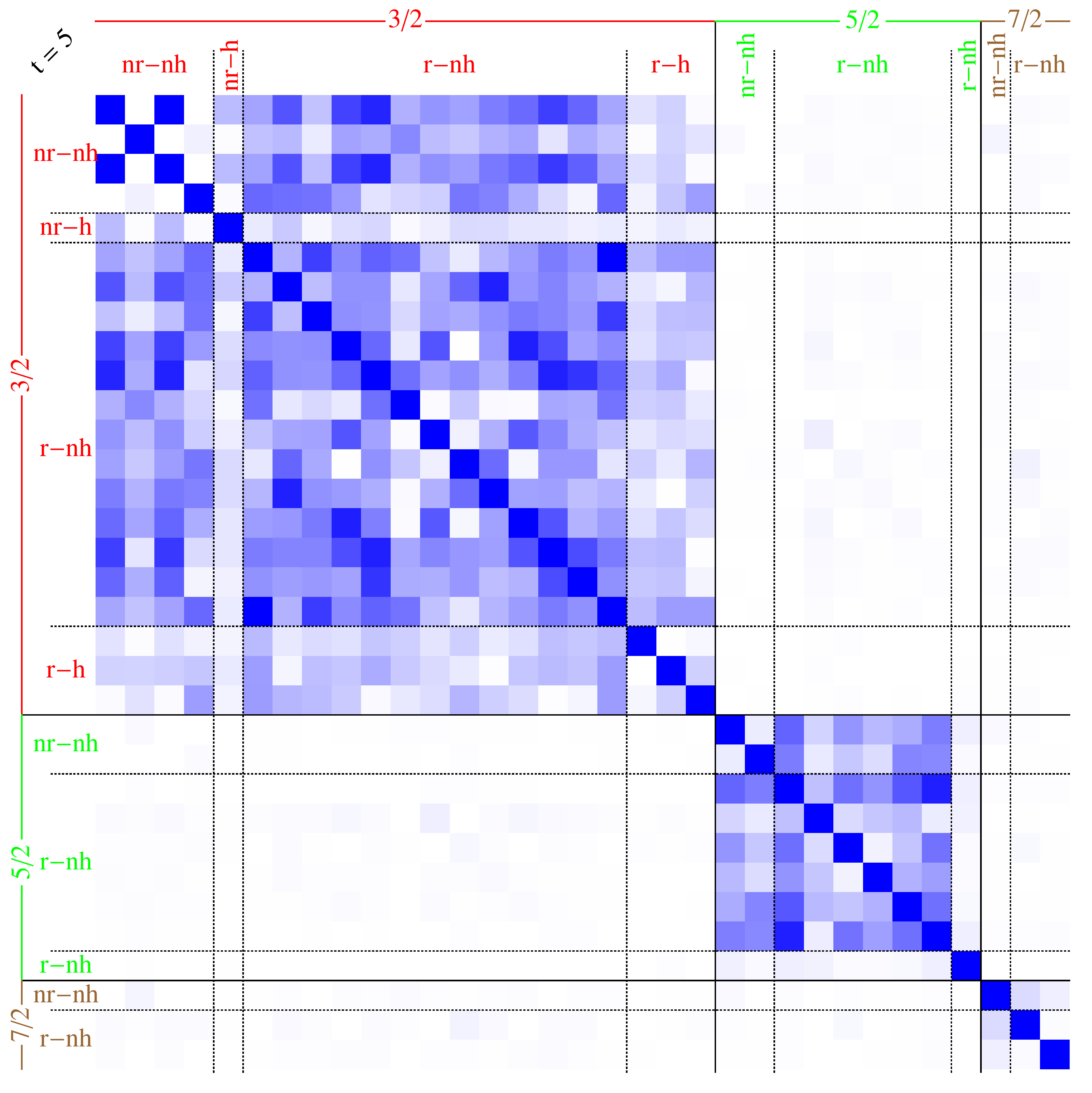}
\includegraphics[scale=0.1]{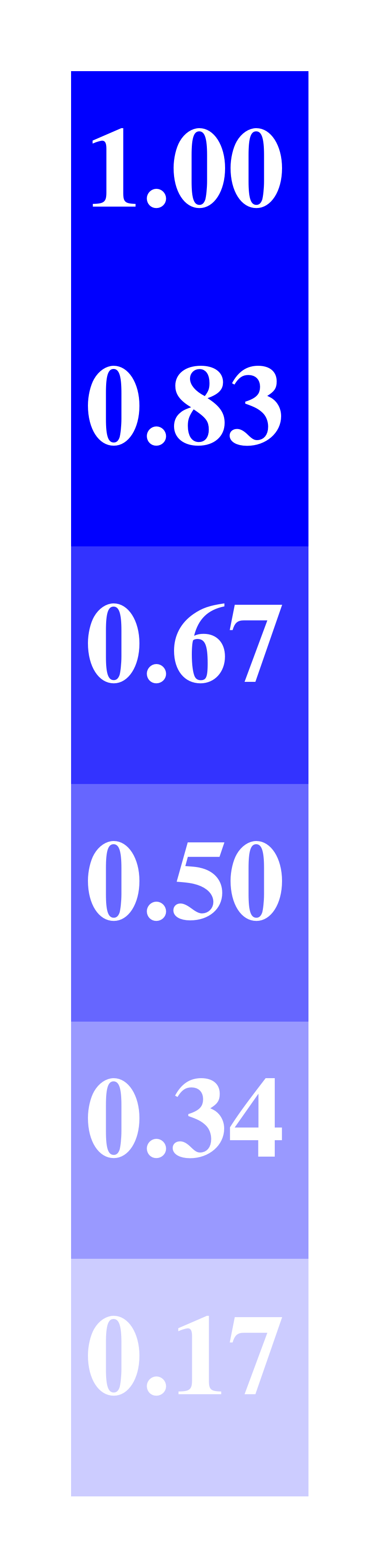}
\caption{A normalized correlation matrix,
  $C_{ij}/\sqrt{C_{ii}C_{jj}}$, at a given time slice at $t/a_t = 5$ for the $H_g$
  irrep.  Operators
  are ordered such that those subduced from spin 3/2 appear first
  followed by spin 5/2 and then spin 7/2. 
}
\eef{rel_Hg_corr_matplot}

\subsection{Rotational symmetry and continuum spin identification} 

One goal of lattice calculations such as this one is to ensure that any states 
identified can be assigned continuum quantum numbers in a reliable way. As 
the continuum limit is taken, rotational symmetry should be restored. The use
of smeared fields in construction operators should to help make this 
restoration more directly observable. 

With the aim of making a more direct link with physical quantum numbers, 
operators were constructed first in the continuum and then subduced onto 
the lattice irreps. Therefore, it is useful to determine whether these 
lattice operators exhibit a remnant of the continuum rotational symmetry on 
the lattice. To check this, \fgn{rel_Hg_corr_matplot} shows the 
correlation function in the $H_g$ irrep at time-separation $5 a_t$ after 
normalizing using ${C}_{ij}/\sqrt{{C}_{ii}{C}_{jj}}$. The choice of 
time-separation here is arbitrary, with alternatives giving similar pictures. 
We do not use this figure for spin identification. 
The normalization ensures all the diagonal entries are unity, while cross 
correlations are always less than 1. Various operators are represented by the 
following abbreviations: `r' indicates relativistic quark spinor components are
used while `nr' indicates non-relativistic spinor only. `h' denotes hybrid
operators and `nh' indicates non-hybrid. There are 33 operators used in this 
irrep, including operators up to two derivatives. The solid lines divide
these operators subduced from continuum spins 
$\frac32, \frac52$ and $\frac72$, and the
dashed lines are used for separating operators defined above with
various abbreviations. The matrix is seen to be almost block diagonal in
the continuum spin label.  A similar pattern is also observed for light and 
strange baryons~\cite{Edwards:2012fx} and our result for triply-charmed 
baryons also suggests that after the subduction a remarkable degree of 
rotational symmetry remains. 
Similar block-diagonal matrices of correlation functions are also
observed within other irreps. 
The approximate block-diagonal structure of these matrices gives confidence in
our ability to make unambiguous spin-identification for triply-charmed baryons.

From this figure, we can also identify the mixing between different 
operators. For example, it is evident from \fgn{rel_Hg_corr_matplot} that 
there is strong mixing between `nr-nh' and `r-nh' type operators and 
comparatively weak mixing between `h' and `nh'-type operators. 
As was observed in Ref.~\cite{Meinel:2012qz}, we also found additional 
suppression of mixing for operators with a given $J$, but with different 
$L$ and $S$, compared to those with the same $J$, as well as the same $L$ and 
$S$. For example, for the operators 
$O_1=\, [(\frac32^+_{1,S}) \otimes D^{[0]}_{L=0,S}]^{\frac32} H_{g1}$
and $O_2=\, [(\frac32^+_{1,S}) \otimes D^{[2]}_{L=0,S}]^{\frac32}H_{g1}$ 
which both have $J = 3/2, L = 0, S = 3/2$ the matrix element 
$M_{12} = C_{12}/\sqrt{C_{11}C_{22}} = 0.99$. 
On the other hand, the mixings of these operators with 
$O_3=[(\frac32^+_{1,S}) \otimes D^{[2]}_{L=2,S}]^{\frac32}H_{g1} $ which
has $J = 3/2, L = 2, S = 3/2$ are $M_{13} = 0.0026$ and $M_{23} =
0.0031$. This suppression of mixing is however evident only for the
non-relativistic operators. We found that for some relativistic
operators mixing enhance between operators with same $J$ but different
$L$ and $S$, compared to those with same $J$, $L$ and $S$.

To identify the spin of a state we followed the same method
detailed in Ref~\cite{Dudek:2007wv} and used in the calculations of
light mesons~\cite{Dudek:2009qf, Dudek:2010wm, Dudek:2010ew}, 
baryons~\cite{Edwards:2011jj, Edwards:2012fx}, charm mesons~\cite{Liu:2012ze}
as well as heavy-light mesons~\cite{Moir:2013ub}. The main ingredient is the 
{\it overlap-factor}, $Z^n_i$, of an operator, $O_i$, to a
state, $n$, which we define as, $Z^n_i \equiv \langle n|O_i^{\dagger}|0\rangle$.
 It is possible to show that these overlap factors enter in the spectral
decomposition of the matrices of the correlation functions as, 
\beq
C_{ij}(t) = \sum_{n} {Z^{n*}_iZ^n_j\over 2 m_n} e^{-m_nt}.  
\eeq{}
Furthermore, one can use the orthogonality for the eigenvectors
$v^{n\dagger}C(t_0)v^{m} = \delta^{n,m}$ to show that the overlap
factors can be obtained from the eigenvectors from the relation 
\beq
Z^n_i = \sqrt{2m_n}e^{m_nt_0/2}v^{n*}_jC_{ji}(t_0).  
\eeq{Z_eq}  
As in previous studies, we utilize these overlap factors for spin 
identification. 
\fgn{gerhist} shows a histogram of these factors for a number of operators onto 
some of the lower-lying states in each of the lattice irreps. The factors 
presented in this figure are normalized according to 
${Z^n_i\over max_n[Z^n_i]}$ such that the largest overlap across all states 
for a given operator is unity. The figure shows clearly that for each state, 
operators coming from a single set subduced from a continuum spin dominate. 
This gives us confidence in the spin assignment for the triply-charmed baryons.
\bef[tbh]
\centering
\includegraphics[width=18cm,height=11cm]{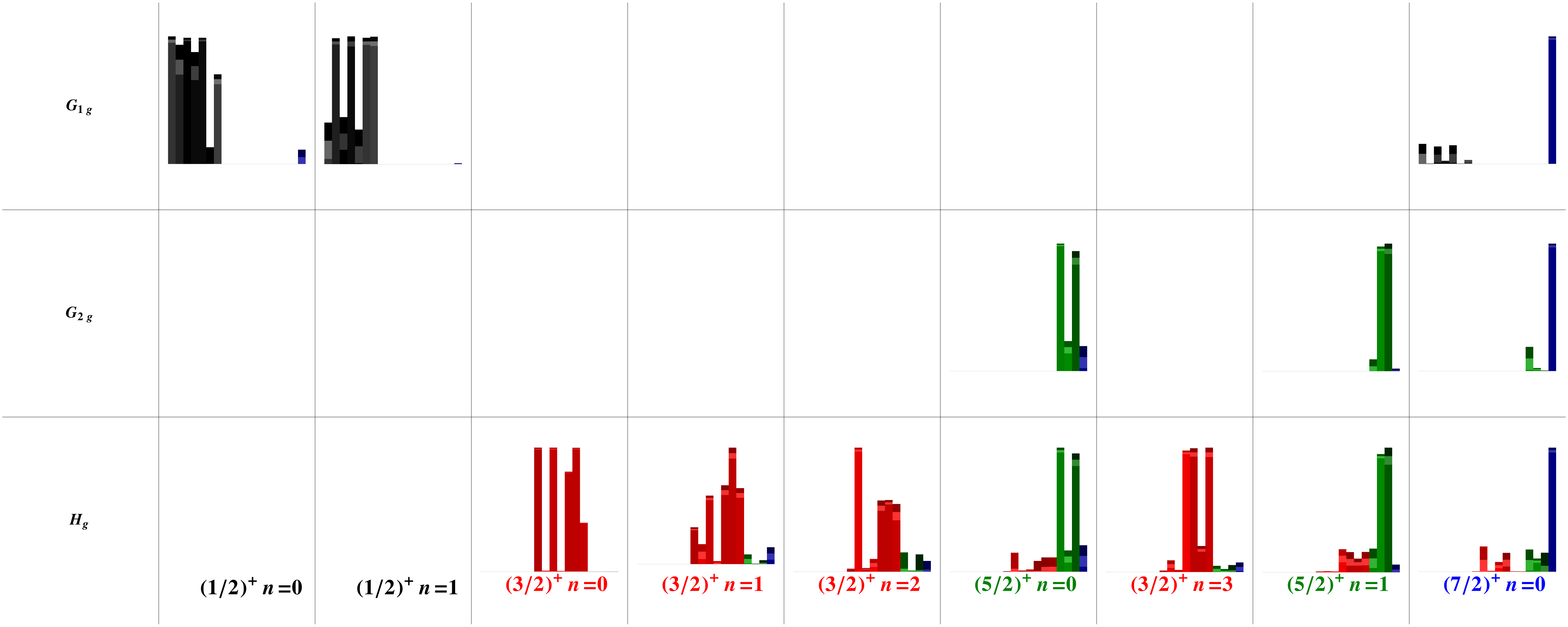}\\
\vspace{-0.2cm}{\large Some of the lower-lying states}\vspace{0.2cm}
\caption{Histograms of the normalized overlap factor, $Z$ of a
  set of operators onto some of the lower-lying states in each 
  lattice irrep. 
The overlaps shown are weighted ${Z^n_i\over max_n[Z^n_i]}$, such 
that the largest normalised value across all states for a given operator is 
unity. 
Various operators are depicted in each state however, due to the large 
basis size, their names are not shown here. 
  Black bars correspond to spin-1/2 state, red for spin-3/2, green for 
  spin-5/2 and blue represents spin-7/2 states. 
  Lighter and darker shades at the top of each bar represent the one-sigma 
  statistical uncertainty.} 
\eef{gerhist}

As in Ref.~\cite{Edwards:2012fx}, one can also show a ``matrix'' plot
to depict the dominant contribution to each state from all
operators. \fgn{rel_G1g_z} and \fgn{rel_Hg_z} show examples for the $G_{1g}$ 
and $H_g$ irreps where the horizontal axis corresponds to the operator index, 
while the vertical location indicates the different states obtained in the
fitting procedure. Solid lines between columns are used to distinguish 
operators subduced from different continuum spins while dashed lines separate 
operators of different types in a given spin.  
\fgn{rel_G1g_z} shows states 2 and 8 predominantly overlap with spin
${7\over 2}^+$ operators and hence can be identified as $J^P =
{7\over 2}^+$ states. For all other fitted states, dominant overlaps
are from spin ${1\over 2}^+$ operators.  However, in order to confirm the
reliability of the identification of a state with a given spin one has
to compare the magnitudes of overlap factors of one operator which is subduced 
into different irreps, which will be discussed later.

These plots give some information on the structure of a state, in particular
the type of operators from which it is constructed. For example, 
states 2 and 8 in $G_{1g}$, which are identified as ${7\over 2}^+$
states, have predominant overlap to non-relativistic non-hybrid and
relativistic non-hybrid operators respectively. Similarly, state 4
overlaps mainly with relativistic non-hybrid operators.  Strong hybrid
content was similarly observed in a number of states.  
It is to be noted from these figures that hybrid operators contribute
predominantly to high-lying states in our observed spectrum. 
\bef[!t]
\centering
\includegraphics[scale=0.3]{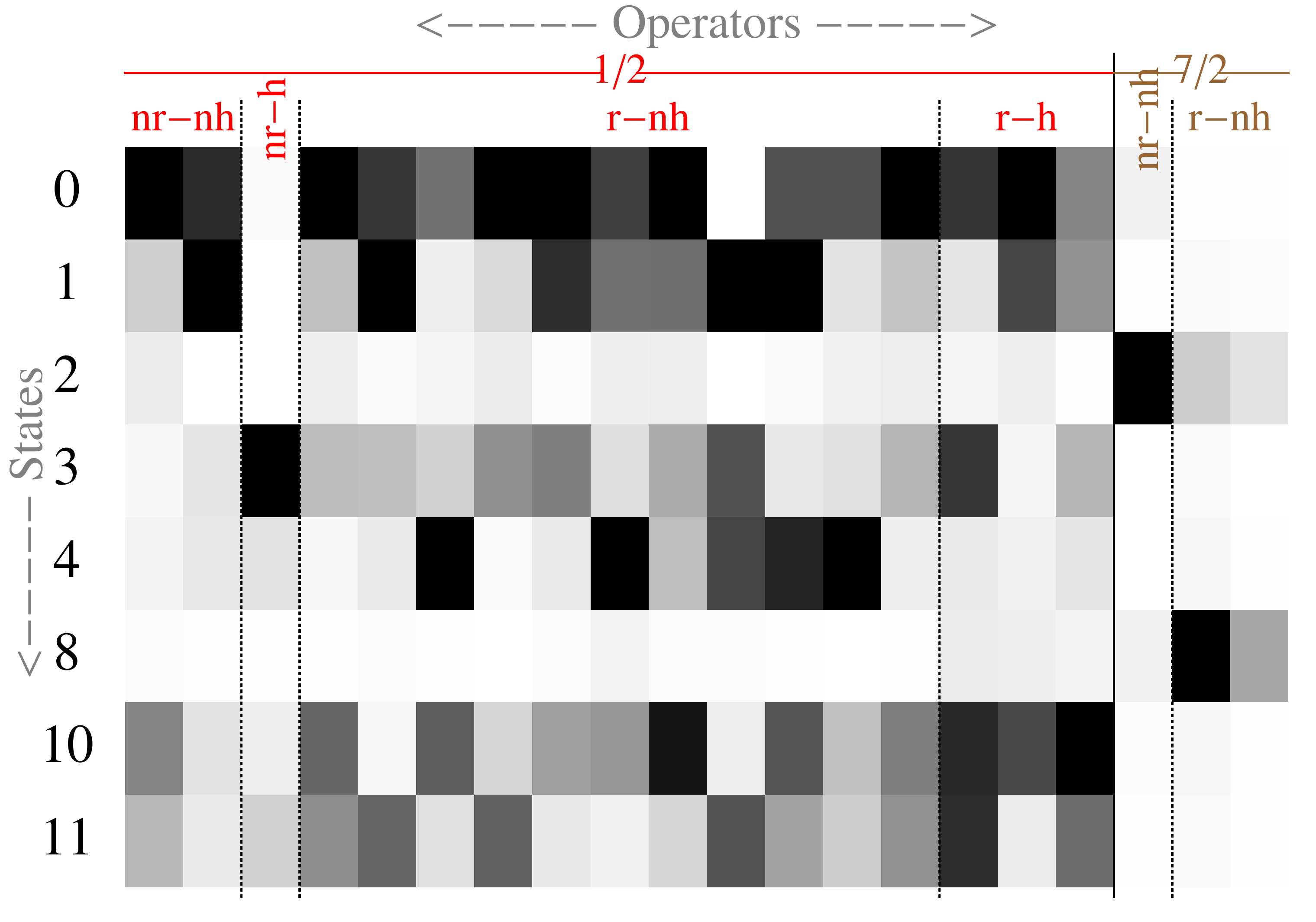}
\includegraphics[scale=0.3]{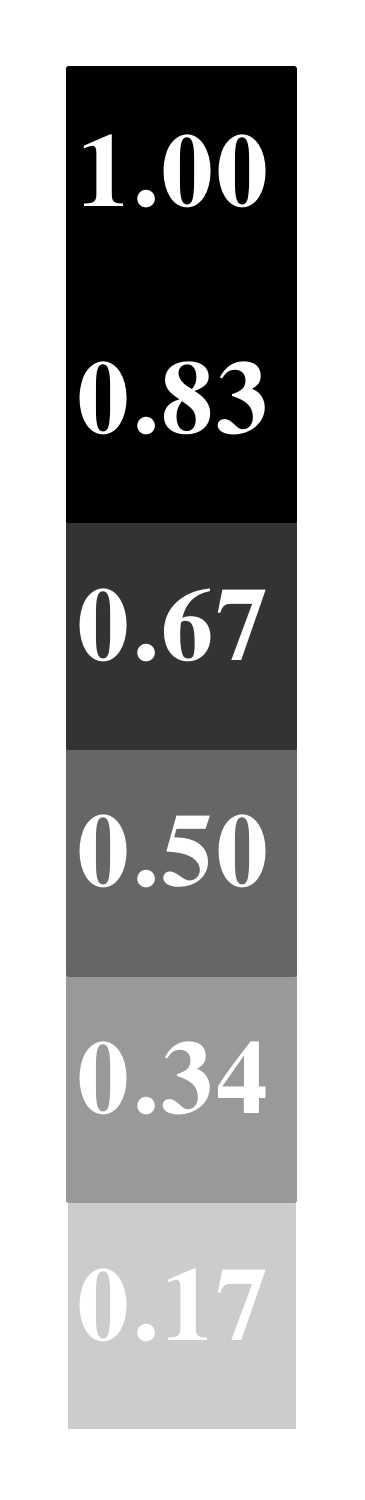}
\caption{``Matrix" plot of values of overlap factor, $Z^n_i$, of an
  operator $i$ to a given state $n$, as defined by \eqn{Z_eq}. $Z^n_i$
  are normalized according to ${Z^n_i\over max_n[Z^n_i]}$, so that for
  a given operator the largest overlap across all states is unity.
  The normalized magnitudes of various operator overlaps to
  states in the $G_{1g}$ irrep are shown. Darker pixel indicates larger 
  values of the operator overlaps
  as shown in the adjacent legend. Various type of operators, for
  example, non-relativistic (nr) and relativistic (r) operators, as
  well as hybrid (h) and non-hybrid (nh) operators are indicated by
  column labels.  In addition, the continuum spins of the operators
  are shown by 1/2 and 7/2. State 0, the ground state, and excited
  states 1, 3, 4, 10 and 11 are identified with $J^P = {1\over 2}^+$
   from the overlap to various types of operators according to
  pixel strength. States 2 and 8 are identified as $J^P = {7\over
    2}^+$ states with the predominant overlap to non-relativistic
  non-hybrid and relativistic non-hybrid operators respectively.}
\eef{rel_G1g_z}
\bef[!b]
\vspace*{-0.2in}
\centering
\includegraphics[scale=0.35]{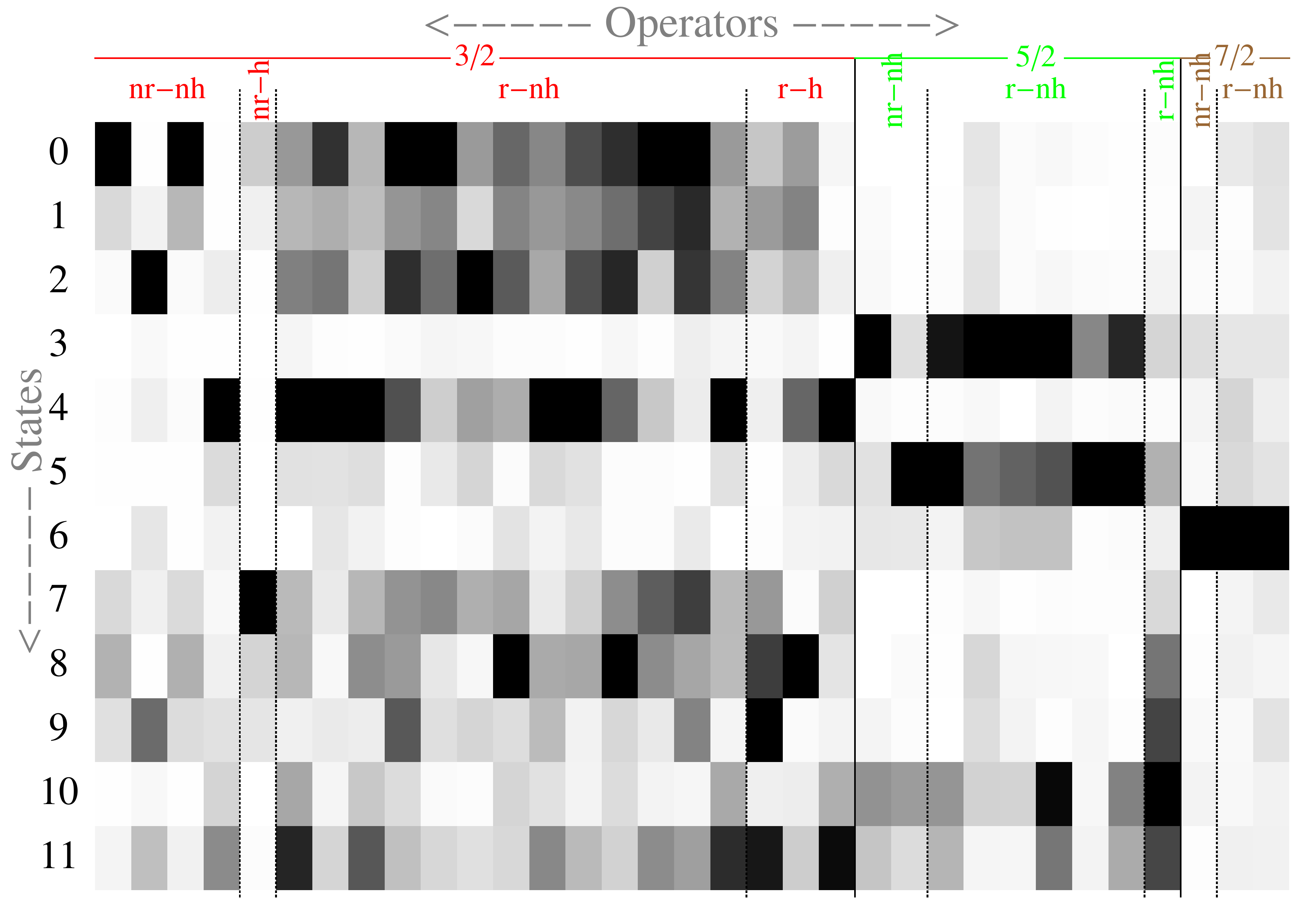}
\caption{Same ``matrix'' plot, as \fgn{rel_G1g_z}, for the $H_g$
  irrep. Here one can associate state 6 with quantum number $J^P =
  {7\over 2}^+$, the states 3, 5 and 10 with $J^P = {5\over 2}^+$, and
  the rest with $J^P = {3\over 2}^+$. The states 7 and 9 are
  predominantly hybrid in nature, while states 4, 8 and 10 are found
  to have substantial overlap with non-hybrid operators. Pixel legend is same as in \fgn{rel_G1g_z}.}
\eef{rel_Hg_z}

To identify a spin contained within a single irrep is relatively easy. For
example, by studying the overlap factors shown in the histogram and matrix 
plots spin-$\frac12$ and spin-$\frac32$ states are clearly identified. However, 
spin-$\frac52$ and spin-$\frac72$ states appear in multiple lattice irreps and 
so more information is crucial. In these cases, an operator can be subduced to 
various irreps but in the continuum limit, overlap factors 
of a continuum operator with a particular state obtained from various cubic
irreps should be almost the same.  
For example, the spin-7/2 continuum operator, 
$[(3/2^{+})_{1,S} \otimes D^{[2]}_{L=2S}]^{\frac72^+}$, can 
be subduced to irrep $G_{1g}$, $H_{g}$ as well as to $G_{2g}$. However,
it is expected that at a finite lattice spacing for a particular
spin $\frac72^{+}$state which is near degenerate over these irreps,
overlap factors of the above operator to that state would also be
degenerate. This near degeneracy of overlap factors identifies this
state as spin-$\frac72^{+}$ state. In \fgn{Z_compare} we compare a
selection of $Z$-values for states conjectured to be $J = \frac52^+$
(top two plots), $\frac52^-$ (middle two plots), $\frac72^+$ (bottom
left) and $\frac72^-$ (bottom right) which can be subduced to
different irreps. 
The values of $Z$ obtained for a given operator from different irreps are 
found to be consistent, or have very small deviations in high-lying states 
which gives confidence in the spin interpretation. Deviations may be due to 
small remaining renormalisation or discretisation artefacts. 

\bef[!h]
\centering
\includegraphics[scale=0.34]{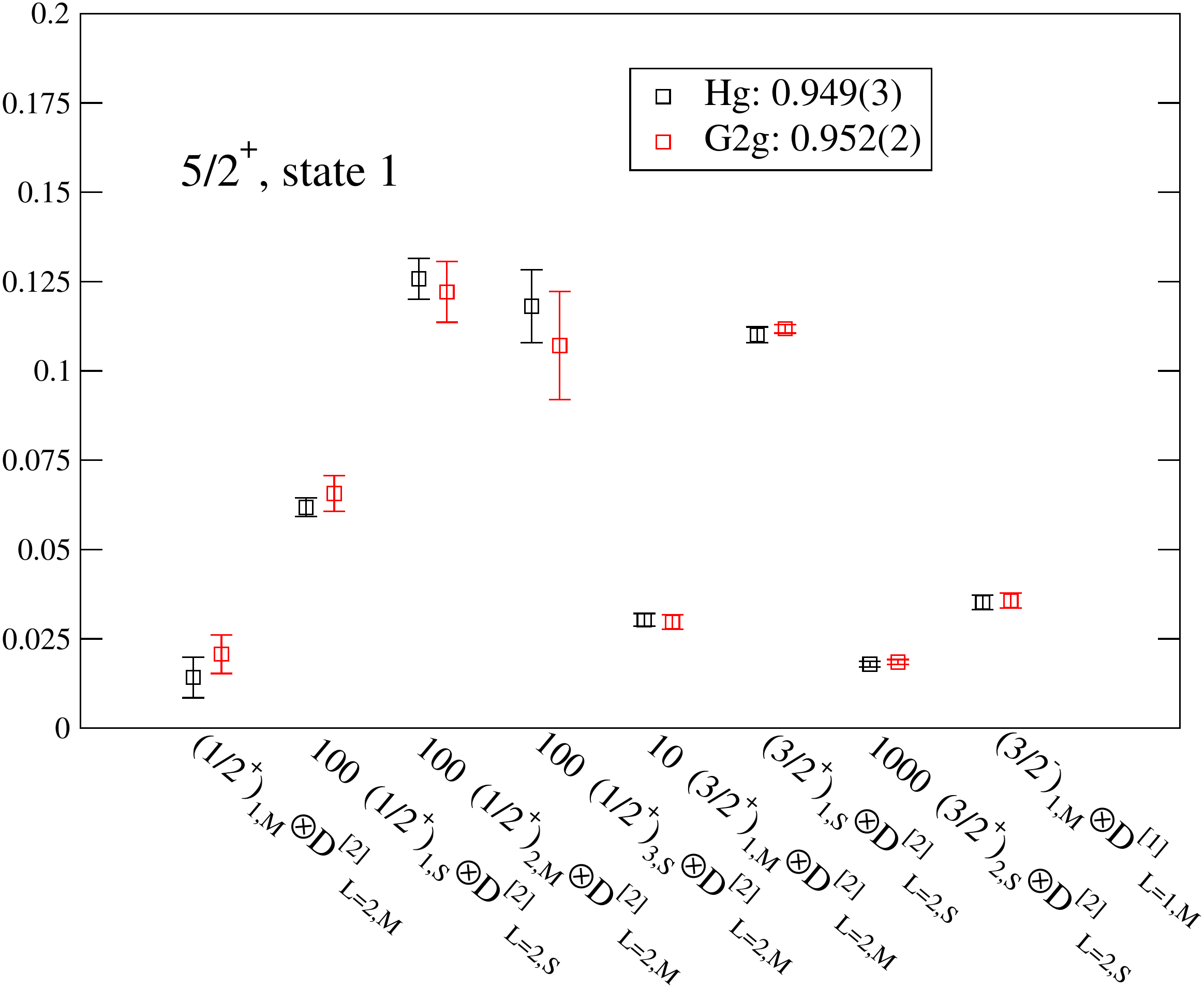}
\hspace*{0.2in}
\includegraphics[scale=0.34]{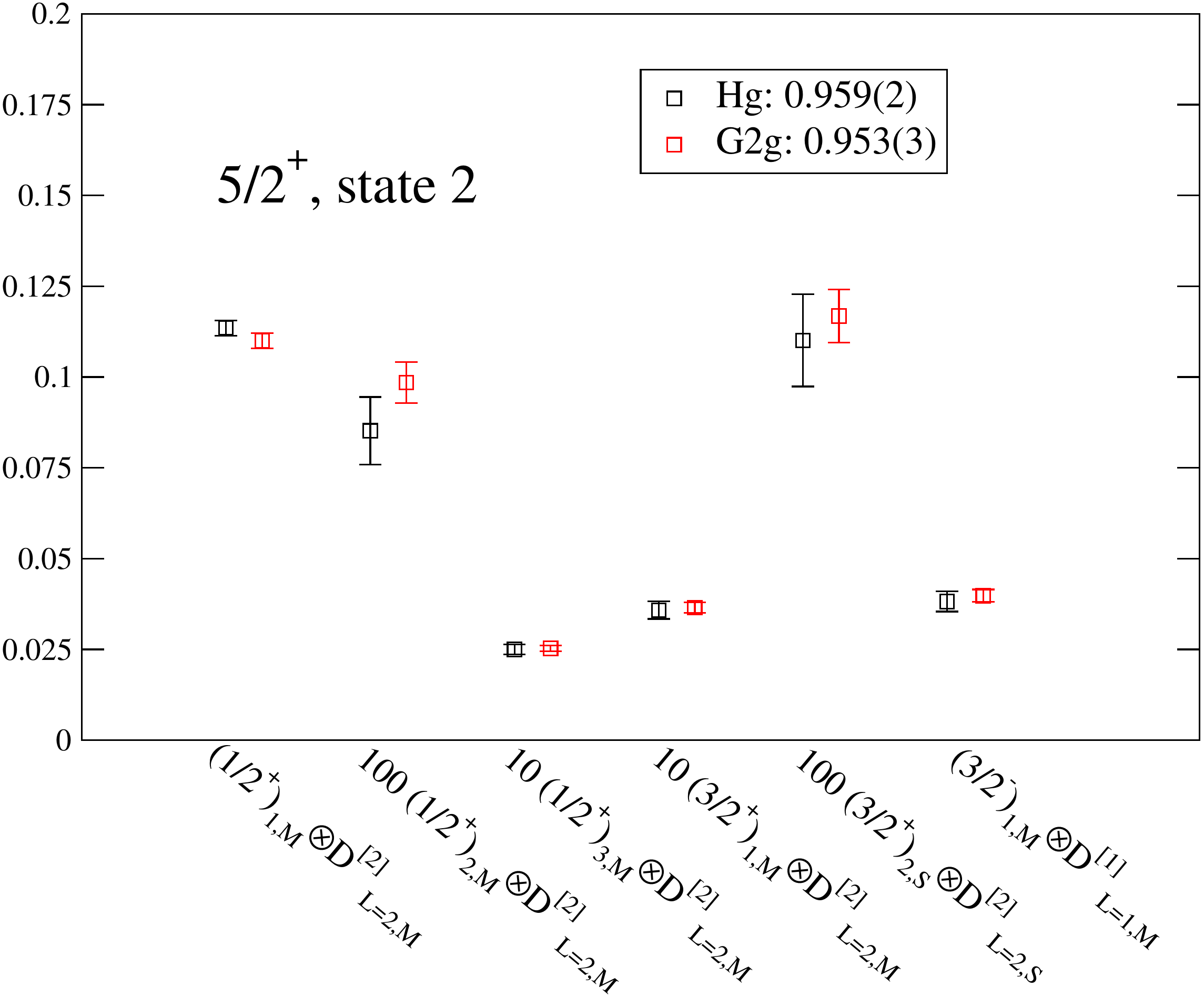}\\
\vspace*{0.3in}
\includegraphics[scale=0.34]{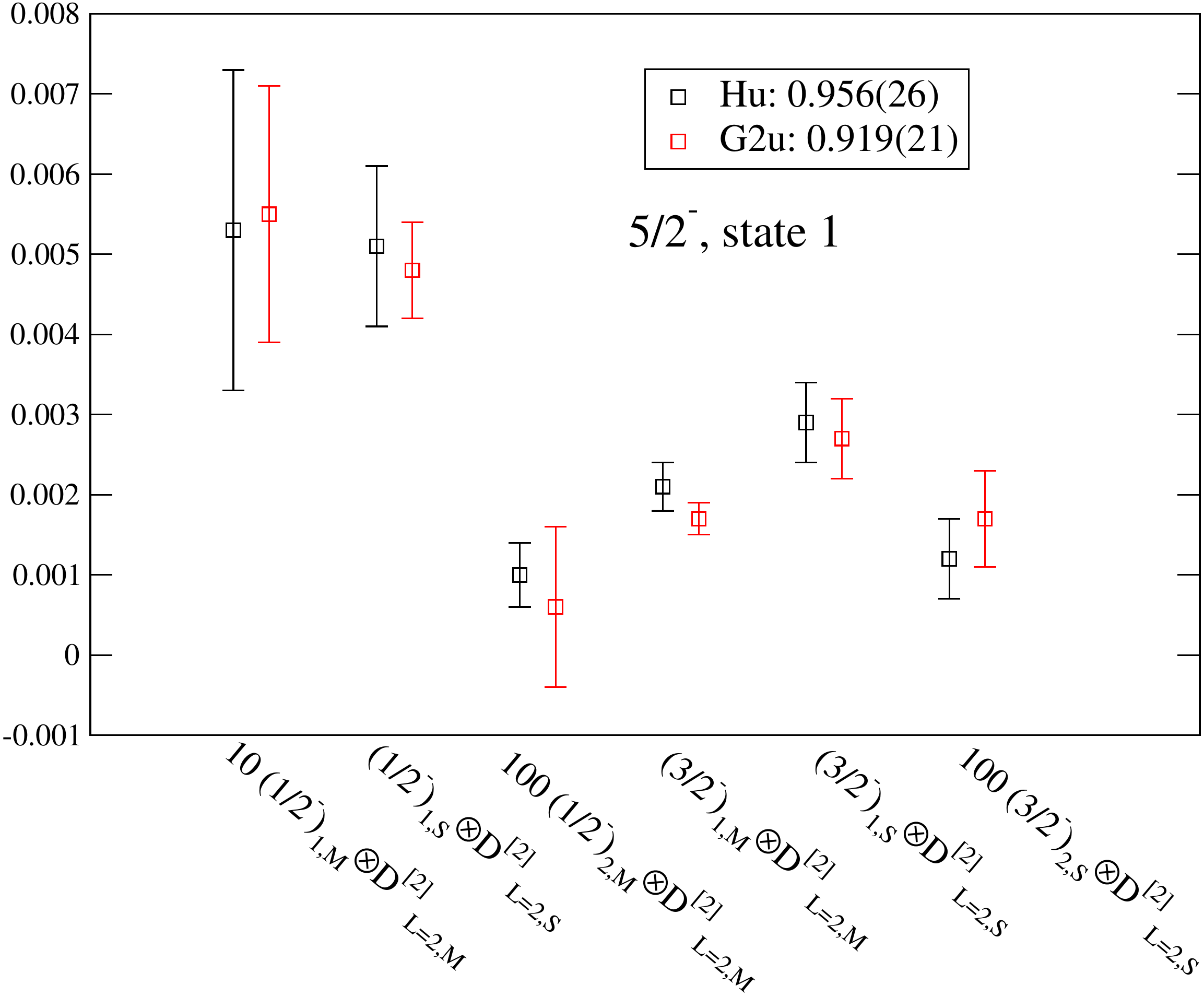}
\hspace*{0.2in}
\includegraphics[scale=0.34]{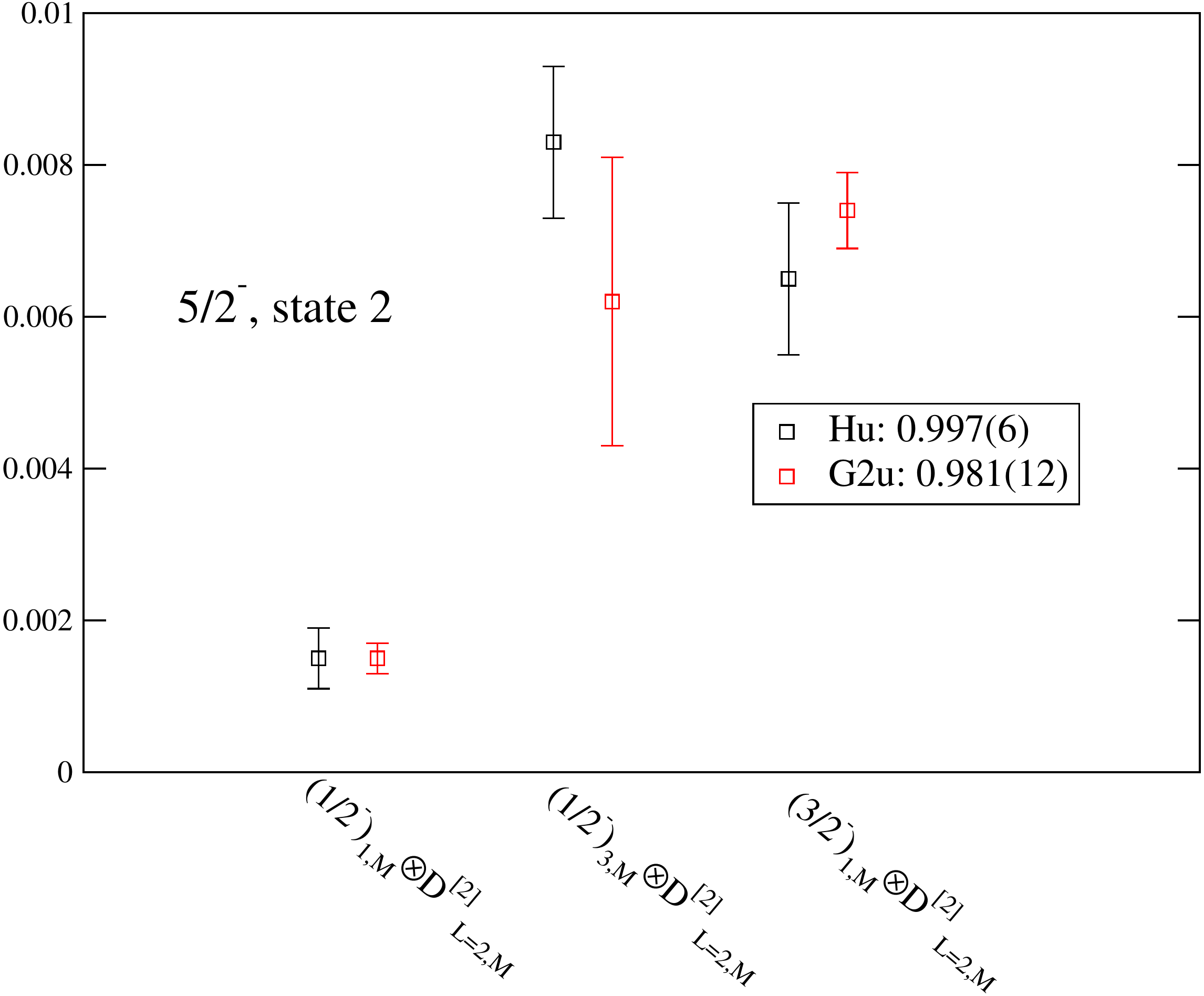}\\
\vspace*{0.3in}
\includegraphics[scale=0.34]{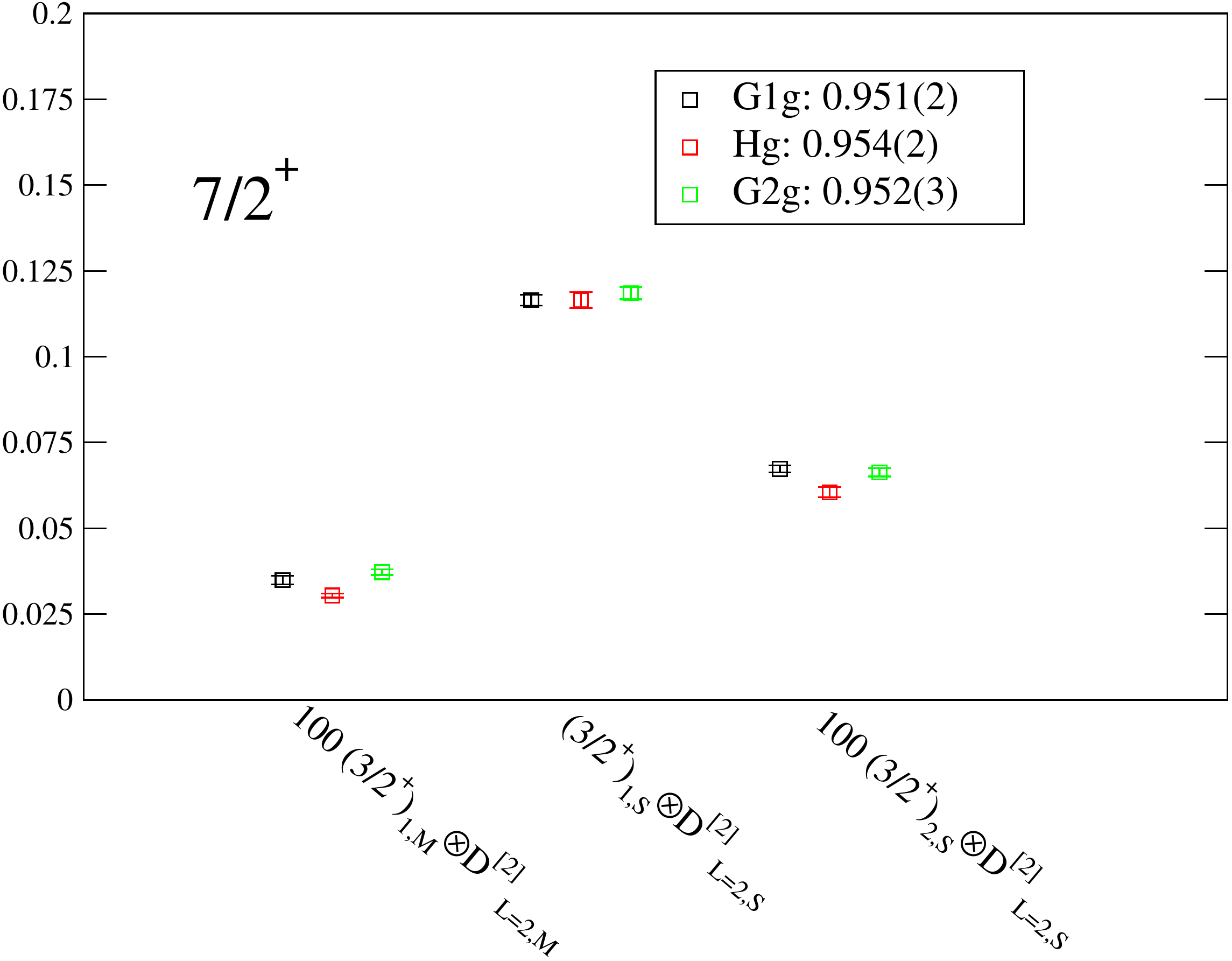}
\hspace*{0.2in}
\includegraphics[scale=0.34]{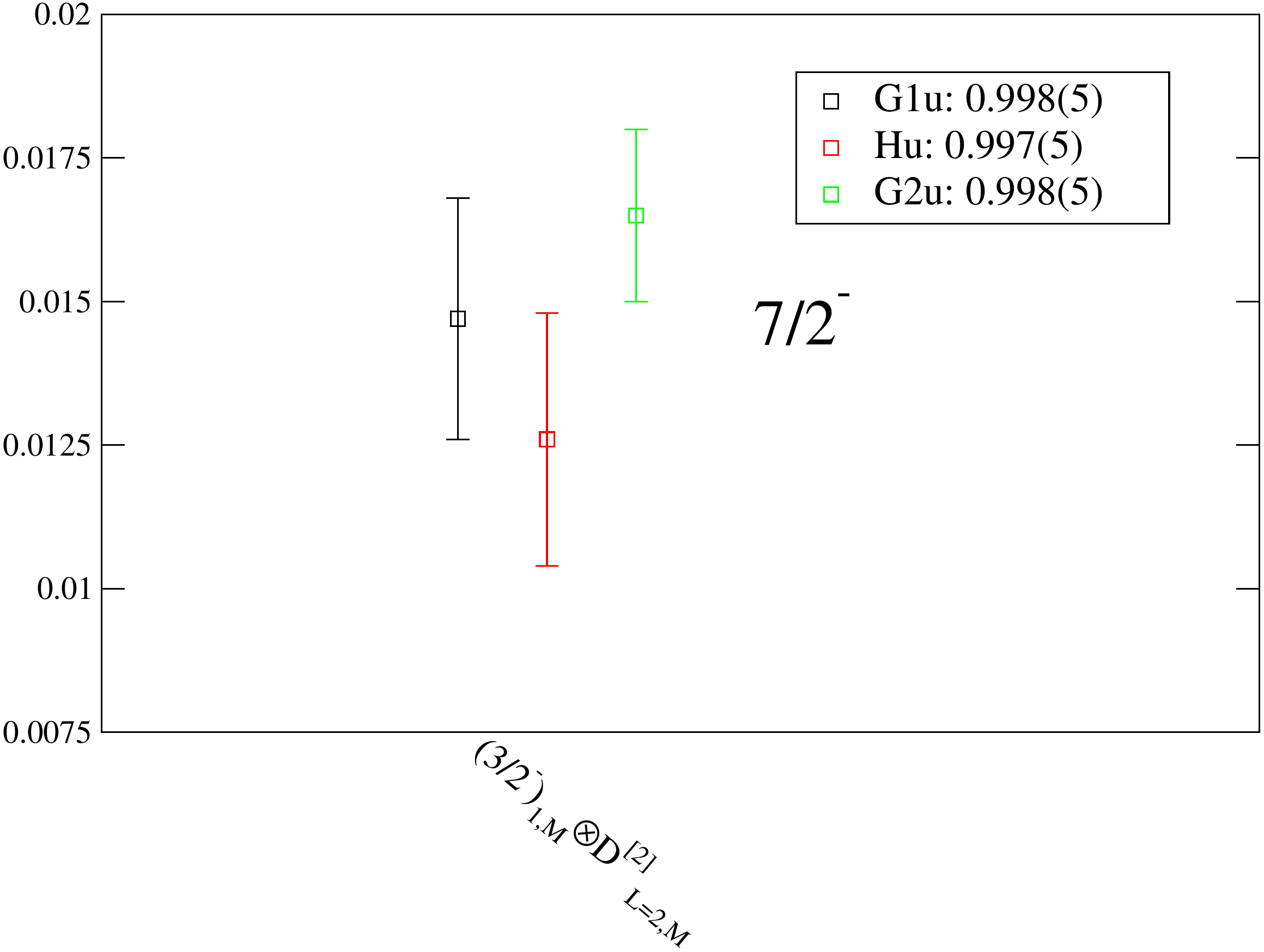}
\caption{A selection of $Z$-values for states conjectured to be
  identified with $J^{P} = \frac52^+$ (top two plots), $\frac52^-$
  (middle two plots), $\frac72^+$ (bottom left) and $\frac72^-$
  (bottom right). Operators in consideration, which overlap to these
  states, are mentioned at the bottom of each plot. $Z$-values
  obtained for a given operator, but from different irreps, are found
  to be consistent or very close to each other which helps to identify the spin of a
  given state. Some operators are scaled so the set can all be shown on a 
single panel. These rescalings are indicated by numbers in front of the 
operator labels. }
\eef{Z_compare}

\subsection{Investigating lattice artefacts}

For a simulation carried out at a finite lattice cut-off, the results obtained 
will differ from their continuum counterparts and the difference 
can be understood in a Symanzik expansion in powers of the lattice spacing. 
Usually, the best means of removing these artefacts is to perform calculations
at a range of lattice spacings and to use the expansion to extrapolate to 
vanishing lattice spacing.  In this study, we do not have sufficient data to 
carry out this extrapolation.
In Ref.~\cite{Liu:2012ze}, a simple experiment was carried out to provide a 
crude estimate of the systematic uncertainty due to ${\cal O}(a)$ 
discretisation artefacts. The charm quark action is discretised using an action
that removes the ${\cal O}(a_s)$ effects at the tree-level. To assess if
radiative corrections to the co-efficient of the improvement term in the 
charm-quark action could lead to significant changes in physical predictions, 
a second calculation was carried out after the co-efficient 
was boosted from the tree-level $c_{s}=1.35$ to $c_{s}=2$. In the charmonium
study, the shift in the mass difference between the two lightest states, the 
$\eta_c$ and $J/\Psi$ was found to be 45 MeV. For the energy difference 
$m_{\Omega_{ccc}}-3/2m_{\eta_c}$, a very similar shift is observed 
in the lowest few states, indicating a similar scale of uncertainty in 
this calculation. The higher lying states do not show any statistically significant difference between the two charm quark actions.  
Here the factor of $\frac32$ corrects for the different number of valence 
charm quarks in a triply-charmed baryon and the $\eta_c$ meson.
\fgn{groundstate_split} shows this alongside other published lattice data, 
which use a different discretisation and so have distinct artefacts. 
Our result for this quantity is consistent with
Refs.~\cite{Basak:2012py, Namekawa:2013vu, Briceno:2012wt,Durr:2012dw},  
but is not consistent with Ref.~\cite{Alexandrou:2012xk}. 
\bef[h!]
\hspace*{-0.3in}
\includegraphics[width=11.cm,height=7cm]{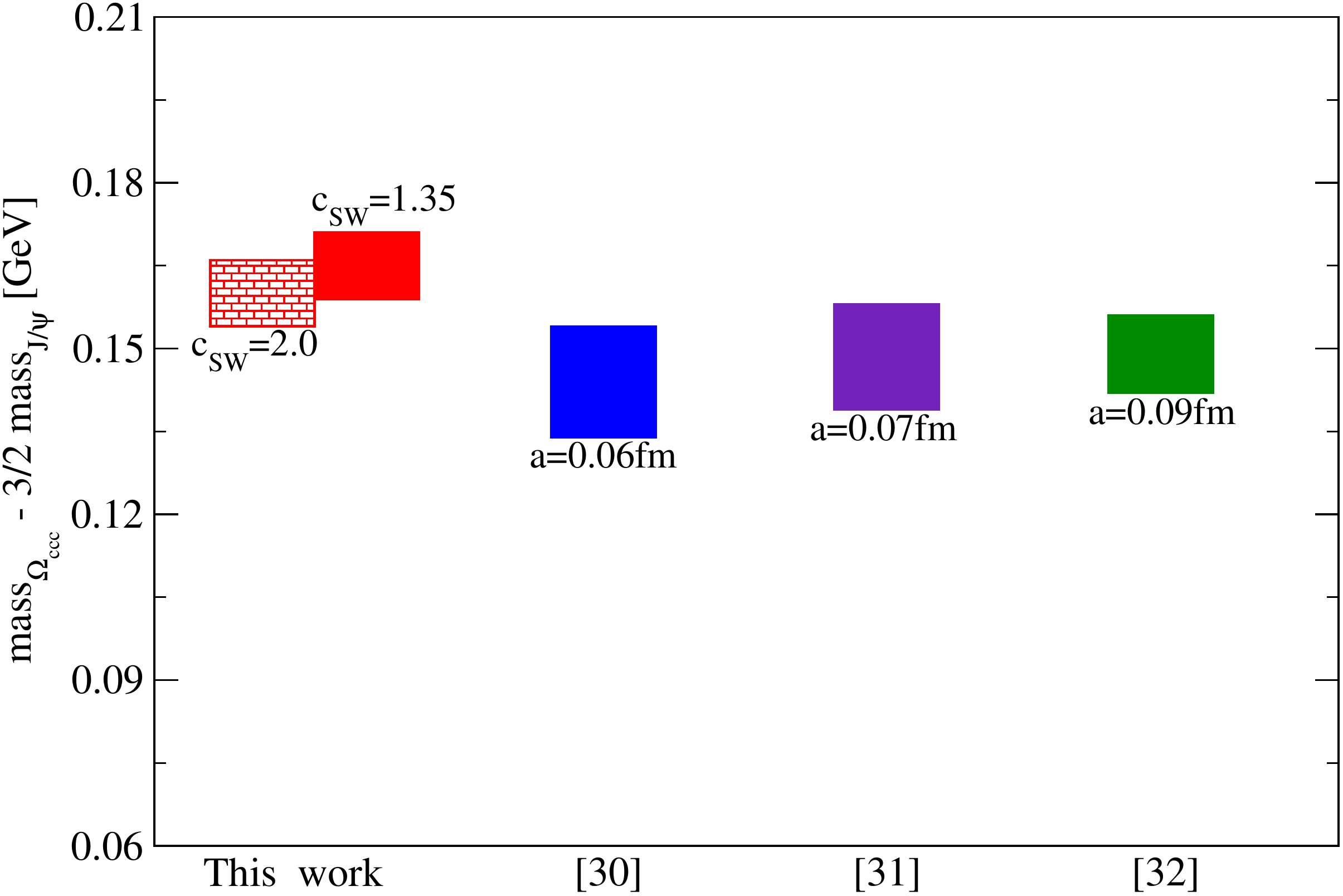}
\caption{Mass splitting of the ground state of $J^{P} = \frac32^+$
  $\Omega_{ccc}$ from ${J/\psi}$ meson. A factor 3/2 is multiplied
  with $J/\psi$ mass to account for the difference in the number of charm quarks  in baryons and mesons. This mass splitting mimics the binding energy of the
  ground state $\Omega_{ccc}$. Results of this mass splitting from
  this work (red boxes) are compared with those obtained from other lattice calculations~\cite{Basak:2012py, Namekawa:2013vu, Briceno:2012wt,Durr:2012dw}.}
\eef{groundstate_split}

\section{Results}

In this section we present our results 
spectra for spins up to $\frac72$ and for both parities. Results for
the extracted masses as well as mass splittings from charmonia states
are presented. Various spin dependent energy splittings
between the extracted states are also considered along with similar
splittings at light, strange as well as bottom quark masses. Results for the
light and strange quark masses are from Ref.~\cite{Edwards:2012fx} 
and bottom quark results are from Ref.~\cite{Meinel:2012qz}.

Due to systematic uncertainty in the determination of the charm quark mass 
parameter in the lattice action, it is preferable to compare energy splittings
between states that reduce the impact of these uncertainties. This also lessens
the effect of ambiguity in the scale setting procedure. In Ref.~\cite{Liu:2012ze},
the quark mass was determined by ensuring the ratio of masses of the $\eta_c$ 
meson and $\Omega$-baryon took its physical value. Consequently for this study,
the spectrum is presented relative to the reference scale of 
$\frac32 m_{\eta_c}$, where the factor of $\frac32$ corrects for the different 
number of valence charm quarks in triply-charmed baryon and charmonium 
states. This subtracted spectrum is presented in 
\fgn{3o2etc_splitting_spectrum_mev} and tabulated in Table \ref{summary_table}. Boxes with thicker borders correspond
to those with a greater overlap onto operators proportional to the field 
strength tensor as discussed in the previous section and which might 
consequently be hybrid states. The states inside the pink ellipses have 
relatively large overlap with non-relativistic operators and should thus be 
well described in a quark model. An analysis of the spectrum using only the 
non-relativistic operators is presented in the right panel of 
\fgn{3o2etc_splitting_spectrum_mev}. The main difference between the two panels
is the absence of spin-5/2 and spin-7/2 negative parity states when only 
operators that couple to the dominant spin components for heavy quarks moving
non-relativistically. 
In Ref.~\cite{Meinel:2012qz}, triply-bottom baryons were studied. The main difference with the set of operators used in that calculation is the inclusion of 
the non-relativistic hybrid operators in this work, which provides access to
higher excited states. 
\bef[!t]
\centering
\includegraphics[scale=0.36]{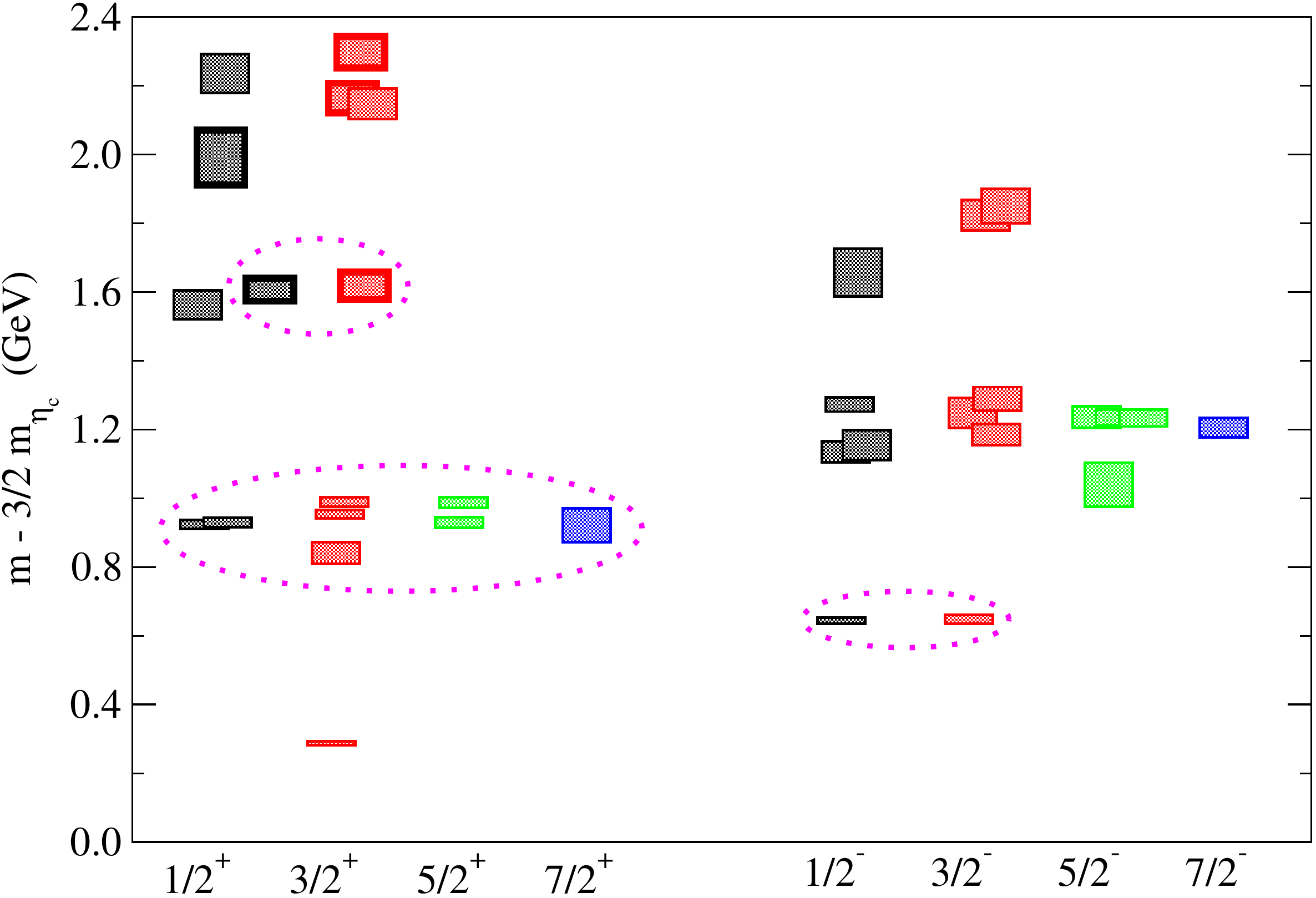}
\includegraphics[scale=0.36]{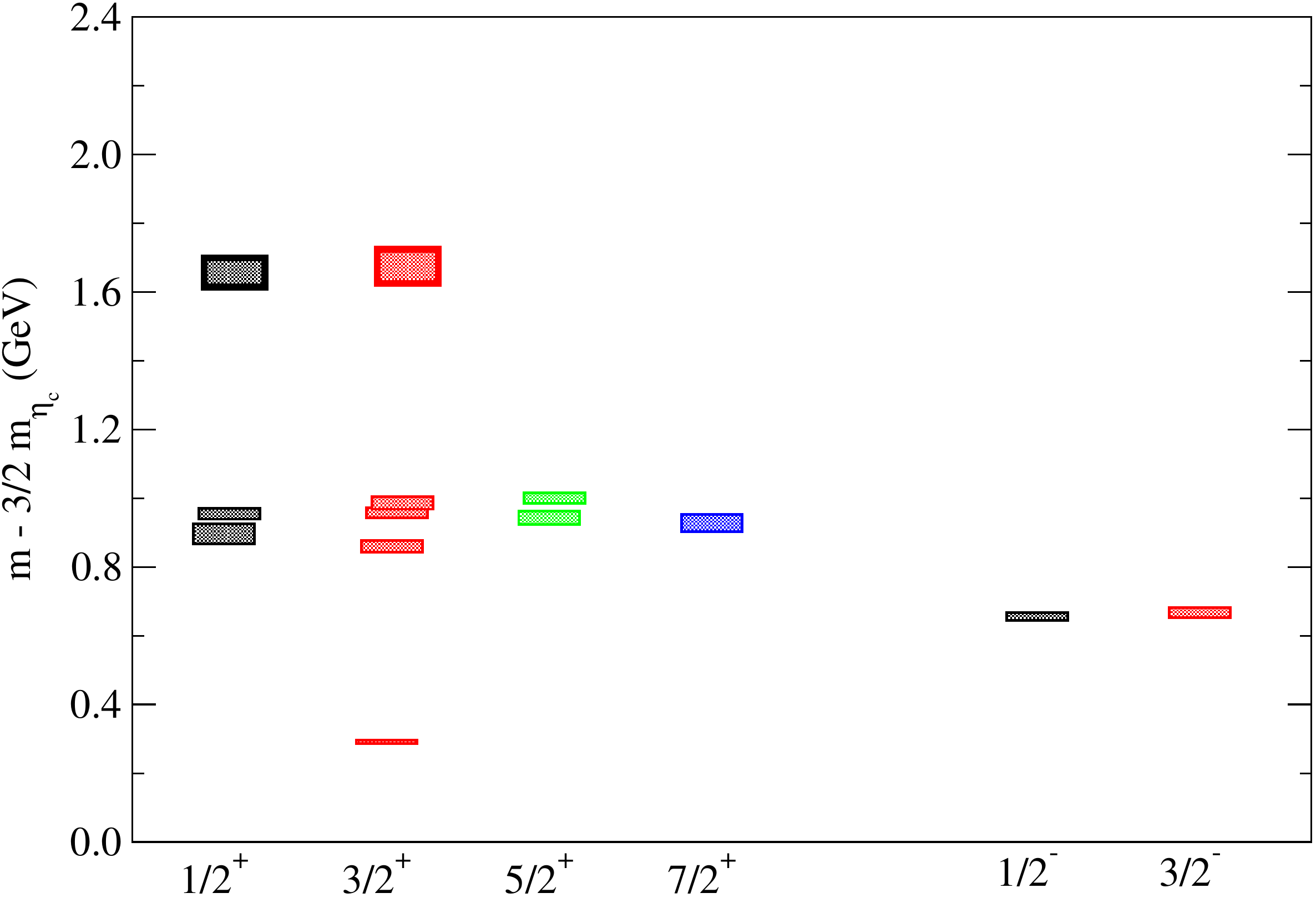}
\caption{Spin identified spectra of triply-charmed baryons with respect to
 $\frac{3}{2} m_{\eta_c}$. 
  The boxes with thick borders
  corresponds to the states with strong overlap with hybrid operators. 
The states inside the pink ellipses are those with relatively large overlap to
  non-relativistic operators.}
\eef{3o2etc_splitting_spectrum_mev}
\begin{table}
\centering
\betb{ c | c | c | c | c | c | c | c }
$\frac12^+$ & $\frac32^+$ & $\frac52^+$ & $\frac72^+$ & $\frac12^-$ & $\frac32^-$ & $\frac52^-$ & $\frac72^-$  \\ \hline \hline
0.923(13)   &  0.287(6)   & 0.930(15)   & 0.921(49)   & 0.644(9)    & 0.648(13)   & 1.040(64)   & 1.205(28)    \\   
0.929(14)   &  0.841(31)  & 0.988(15)   &             & 1.136(31)   & 1.186(31)   & 1.233(25)   &              \\         
1.563(33)   &  0.954(13)  &             &             & 1.155(43)   & 1.248(44)   & 1.234(24)   &              \\         
1.607(42)   &  0.989(13)  &             &             & 1.273(21)   & 1.289(34)   & 1.236(32)   &              \\         
1.992(80)   &  1.618(40)  &             &             & 1.656(69)   & 1.823(44)   &             &              \\         
2.236(56)   &  2.147(46)  &             &             &             & 1.850(51)   &             &              \\         
            &  2.165(43)  &             &             &             &             &             &              \\         
            &  2.298(46)  &             &             &             &             &             &                         
\eetb
\caption{Spin identified spectra of triply-charmed baryons with respect to
 $\frac{3}{2} m_{\eta_c}$ (that is $m_{_{\Omega_{ccc}}} - \frac32~m_{\eta_c}$.)}
\label{summary_table}
\end{table}

In the lowest positive-parity band and the two lowest negative-parity bands the
number of states for each spin agrees with the expectation given in 
\tbn{n_operators}. In that table, the number of states is presented for 
operator built from up to two derivatives and this corresponds to the number 
allowed by $SU(6)\times O(3)$ symmetry. 
In this picture for $N_D=0$, only one state with $J^P=\frac32^+$
is expected. For $N_D=1$ the expected quantum numbers are $J^P=\frac12^-$ and
$J^P=\frac32-$ and for $N_D=2$, a set of positive parity states with a range 
of spins from $\frac12$ to $\frac52$ is predicted with multiple states 
appearing for the lower spins. This pattern is clearly seen in the spectrum 
determined by this calculation.
This agreement of the number of low lying states between the lattice
spectra obtained in this work and the expectations based on
non-relativistic quark spins implies a clear signature of $SU(6)\times
O(3)$ symmetry in the spectra. Such $SU(6)\times O(3)$ symmetric
nature of spectra was also observed in Ref.~\cite{Edwards:2012fx}.  As
was pointed out in Ref.~\cite{Edwards:2012fx}, it is not meaningful to
interpret the higher excited states in terms of $SU(6)\times O(3)$
symmetry. The reasons behind this are following : firstly, we did not
include non-relativistic operators with three derivatives and secondly
for higher excited states it is expected that the relativistic
operators generally overlaps more with these states. For negative parity
it is also not possible to identify a state
with strong hybrid content because it is not clear how the relative
importance of all the relevant operators overlapping to that state
will change in the presence of non-relativistic operators having three
or more number of derivatives.

\bef[tbh]
\centering
\includegraphics[scale=0.5]{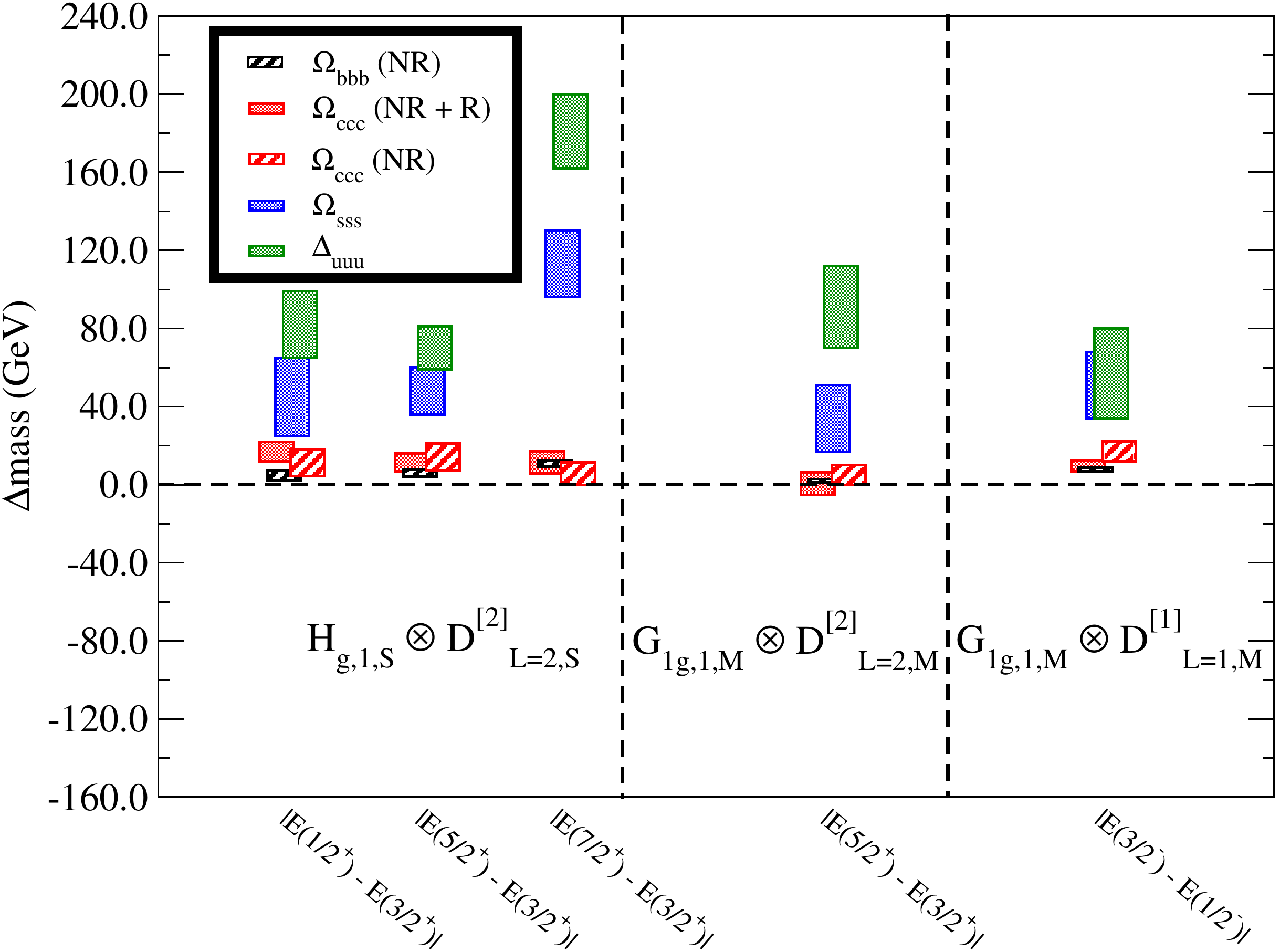}
\caption{Energy splittings between states with same $L$ and $S$ values,
  starting from light to heavy baryons. For $\Omega_{bbb}$, results
  are with only non-relativistic operators~\cite{Meinel:2012qz}; for
  $\Omega_{ccc}$, results from relativistic and non-relativistic as
  well as only non-relativistic operators are shown, and for light and
  strange baryons results are with relativistic and non-relativistic
  operators~\cite{Edwards:2012fx}. These results are obtained from
  fitting the jackknife ratio of the correlators which helps to get
  smaller errorbar in splittings. The left column is for the states
  with $D = 2, S = \frac32$ and $L = 2$. The symbol $H_{g,1,S}$ refers
  to the first embedding of irrep $H_g$ in the totally symmetric Dirac
  spin combination, while $D^{[2]}_{L=2,S}$ refers to spatial
  projection operators with two derivatives in a totally symmetric
  combination, and with orbital angular momentum two. Similarly, the
  middle column is for the states with $D = 2, S = \frac12$ and $L =
  2$. Here irrep is $G_{1g}$ and both Dirac and derivative are in a
  mixed symmetric combination. In the right column these negative
  parity states have $D = 1, S = \frac12$ and $L = 1$. Here again
  irrep is $G_{1g}$ and both Dirac and derivative are in a mixed
  symmetric combination.}
\eef{split_ccu_n_comp}

\subsection{Valence quark mass dependence of energy splittings}

Spin-dependent splittings between the triply-flavoured baryons provide an 
insight into interactions between three confined quarks with the same mass. We 
consider now the dependence of these splittings on the mass of the valence  
quarks. 
\bef[tbh]
\centering
\includegraphics[scale=0.32]{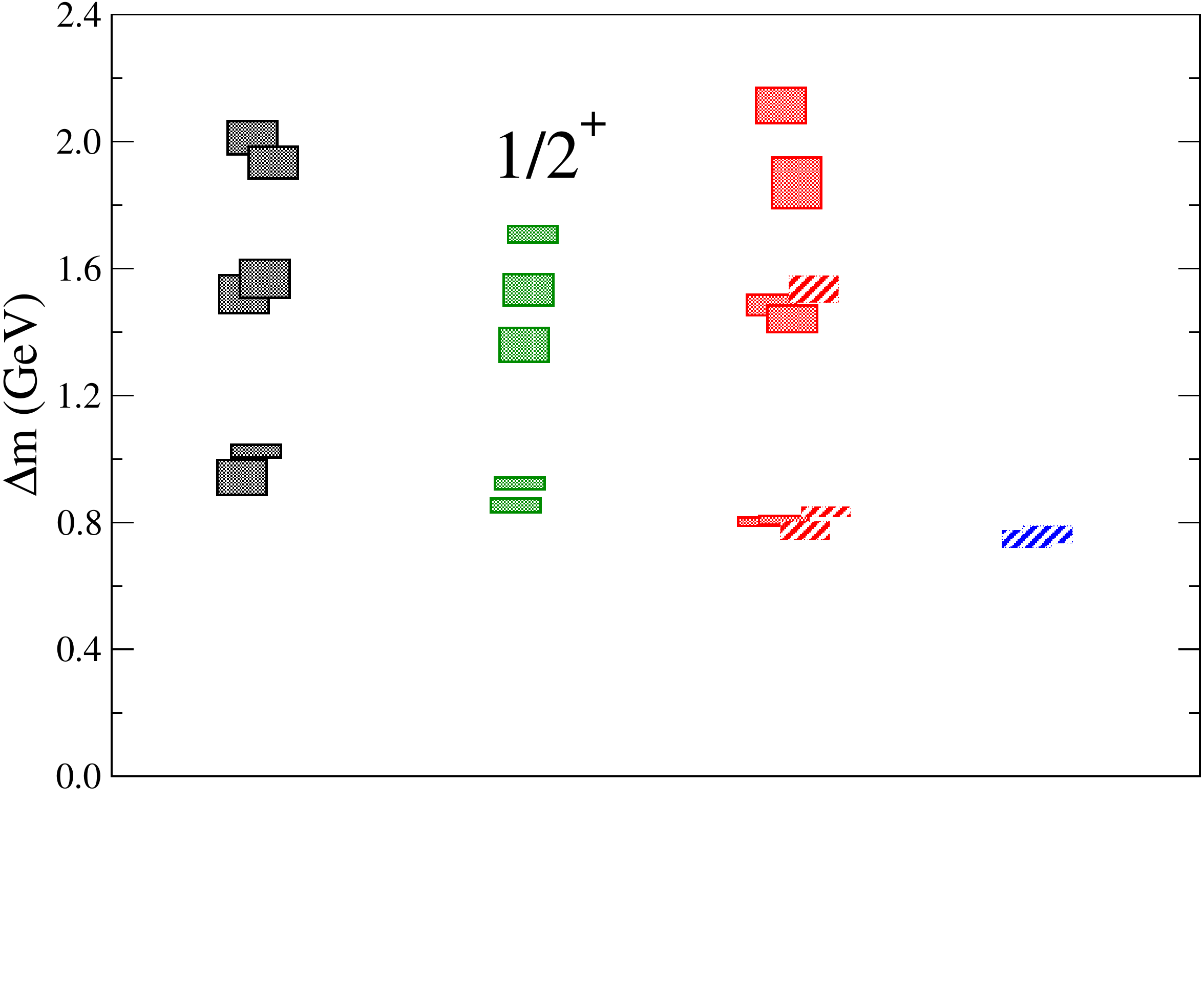}
\hspace*{0.15in}
\includegraphics[scale=0.32]{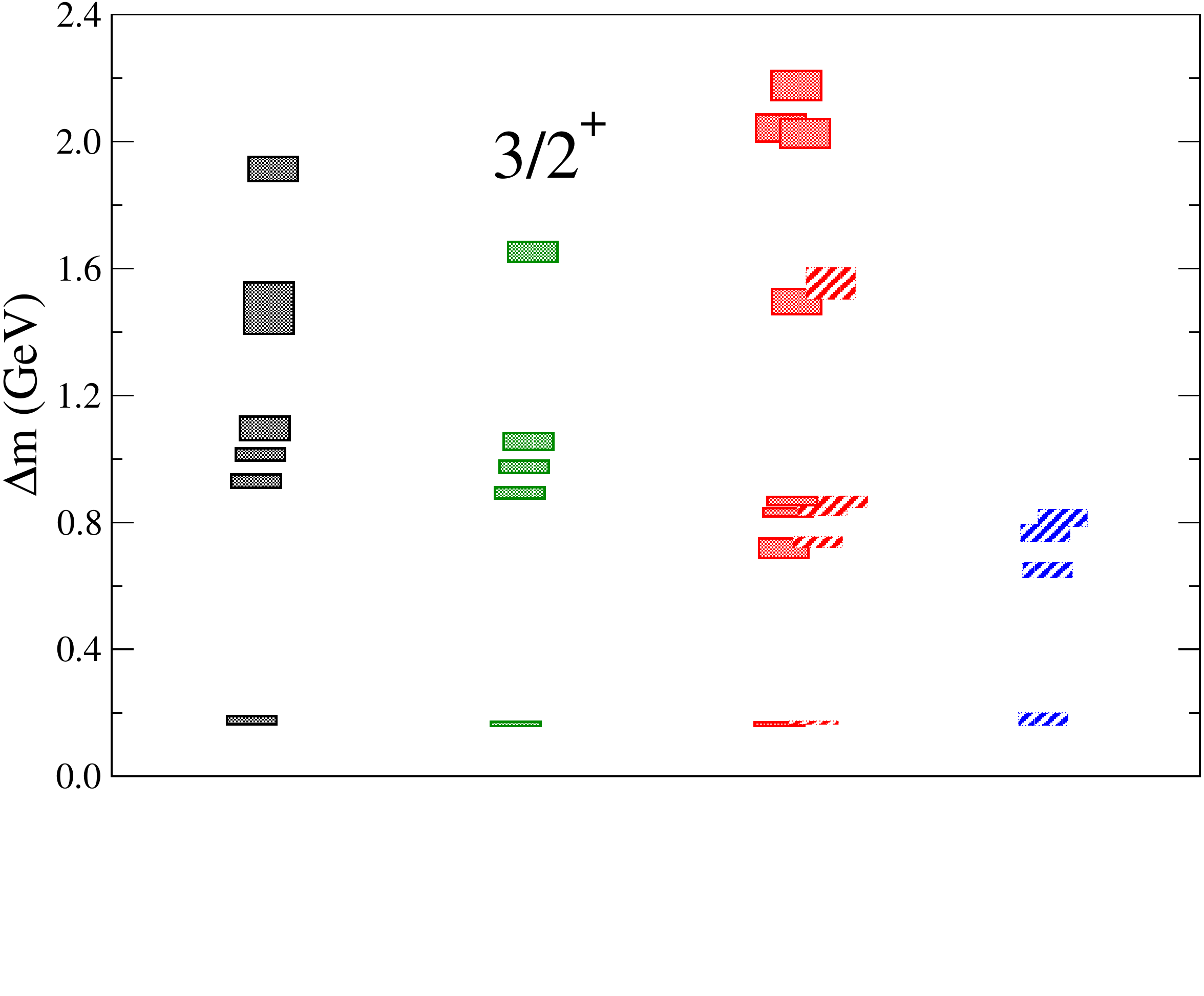}\\
\vspace*{-0.45in}
\includegraphics[scale=0.32]{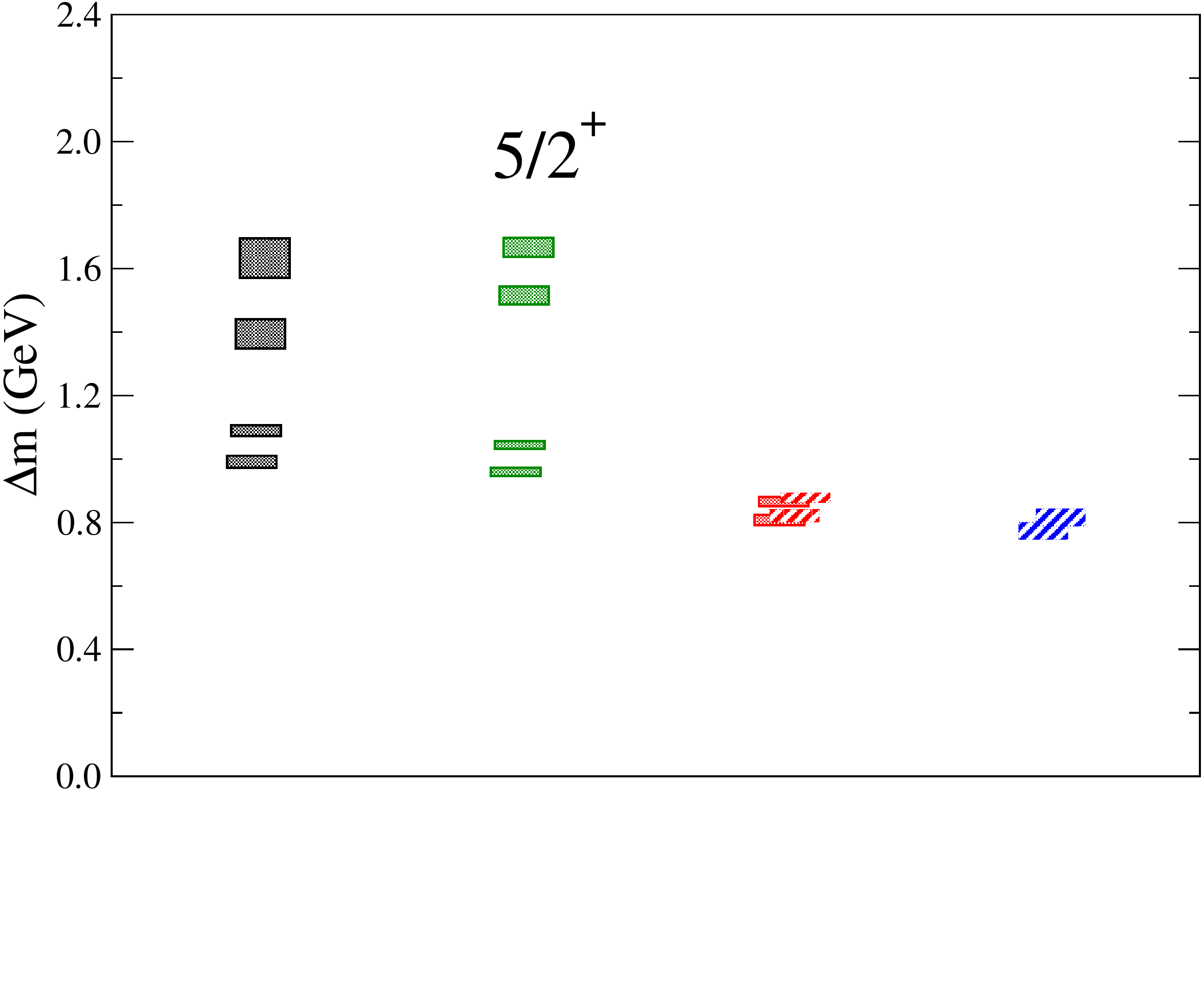}
\hspace*{0.15in}
\includegraphics[scale=0.32]{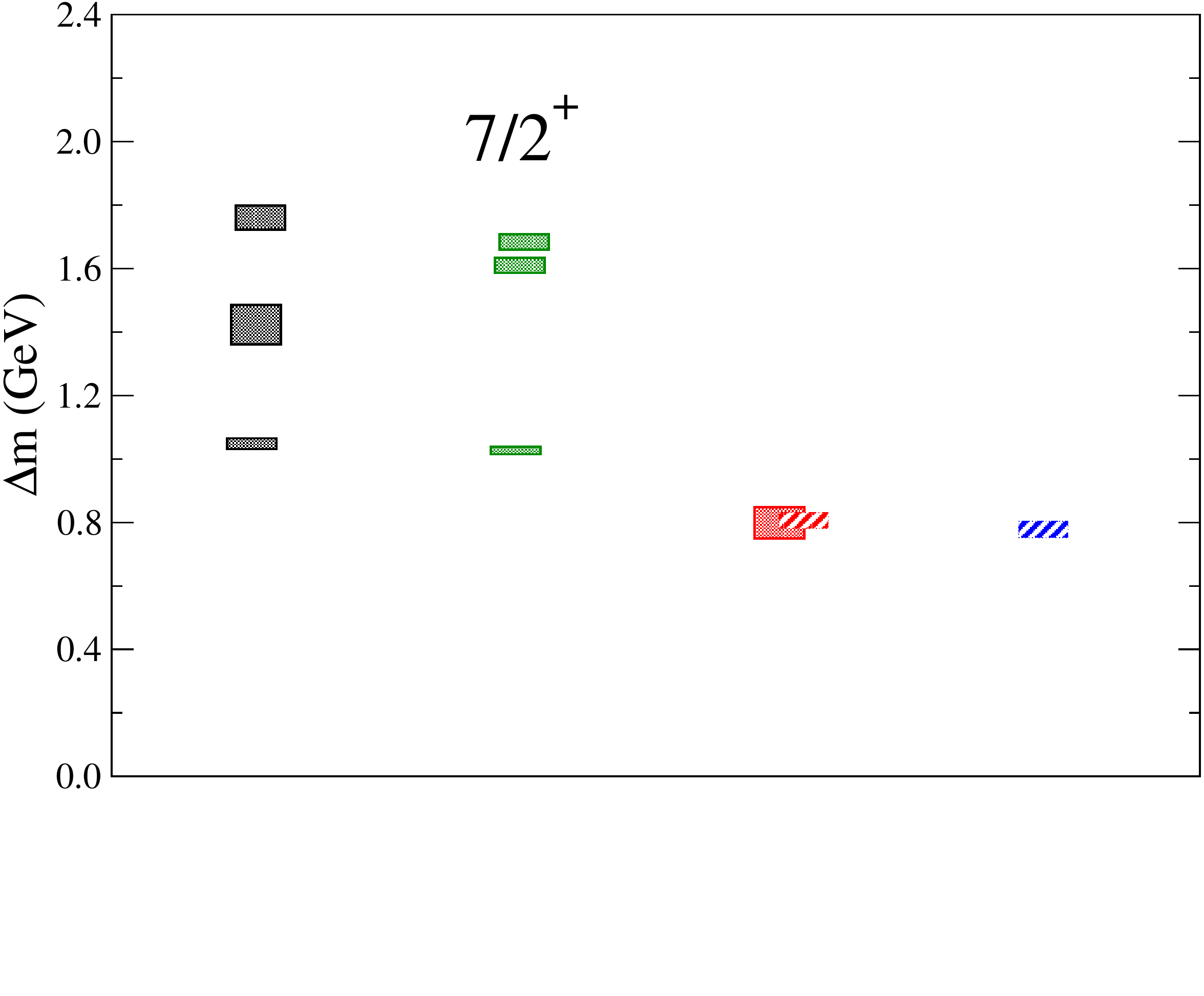}\\
\vspace*{-0.45in}
\includegraphics[scale=0.32]{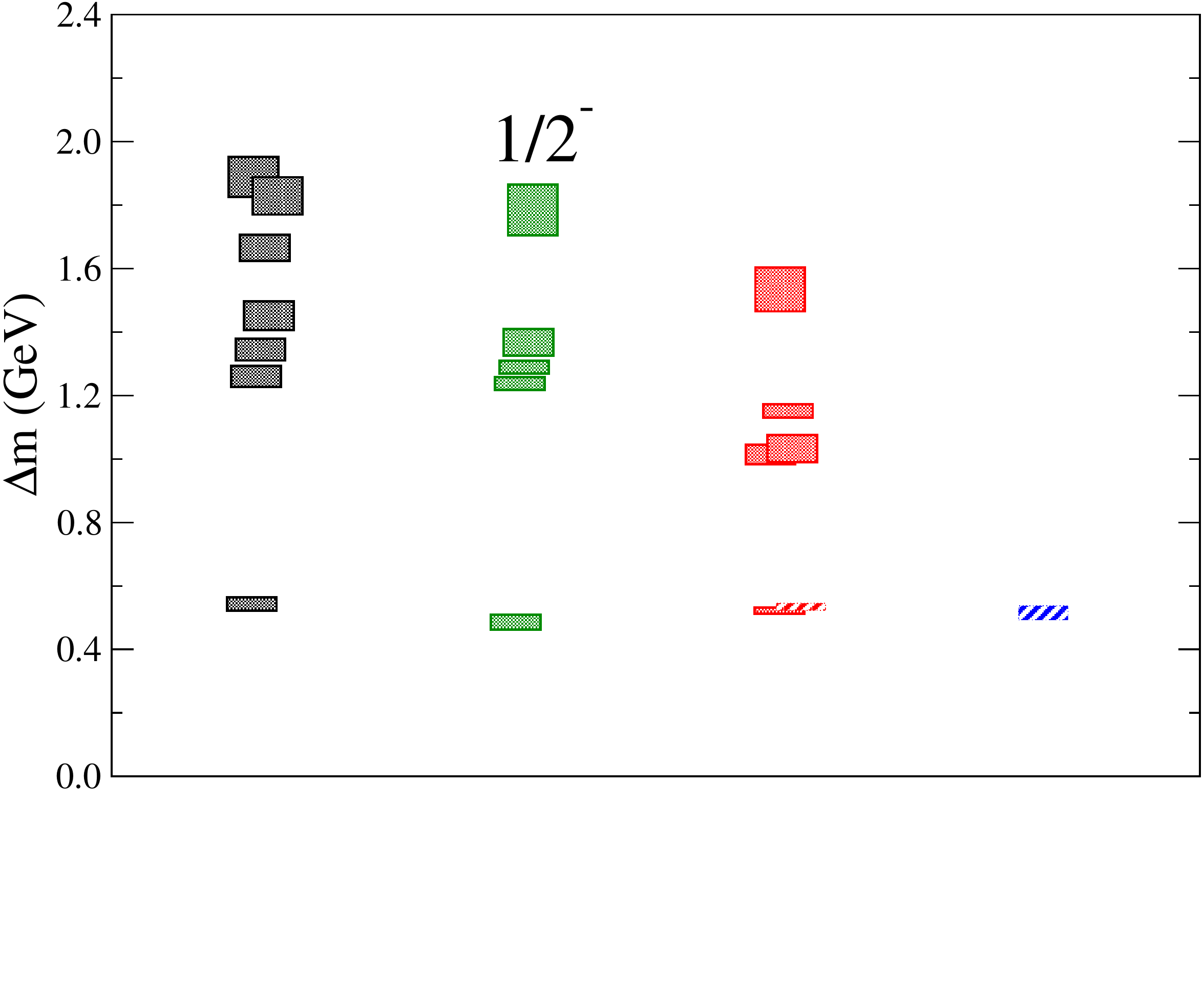}
\hspace*{0.15in}
\includegraphics[scale=0.32]{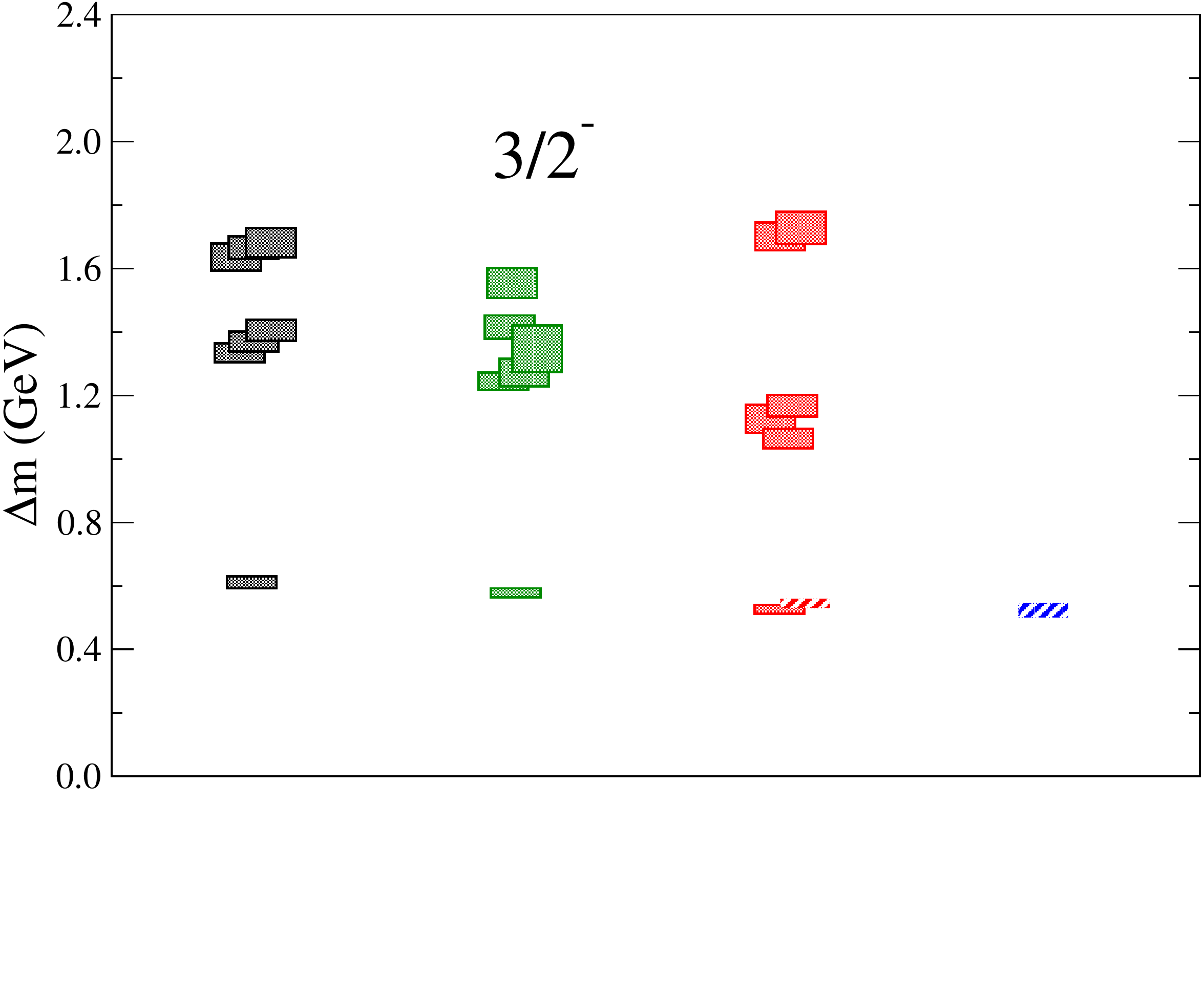}\\
\vspace*{-0.45in}
\includegraphics[scale=0.32]{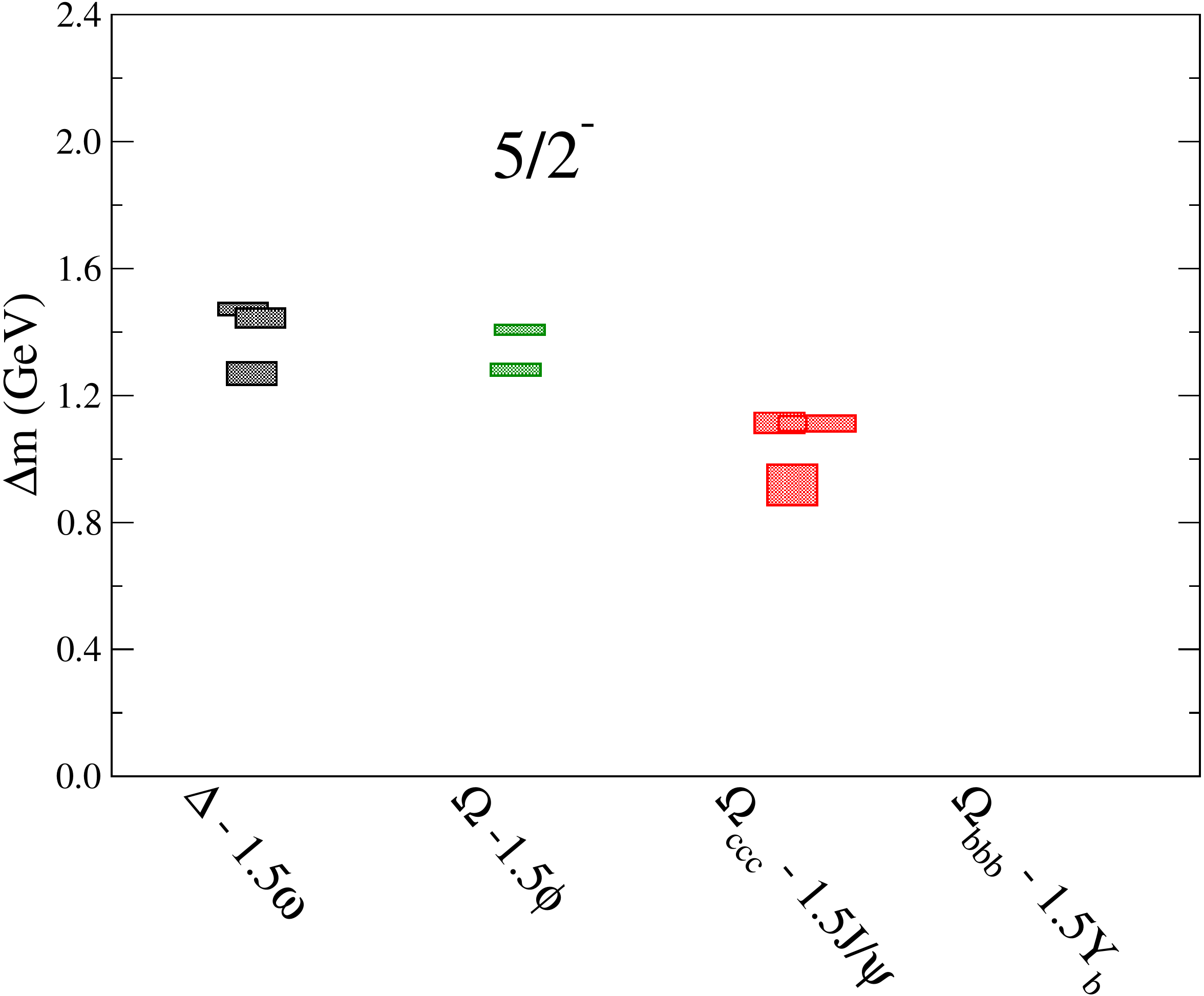}
\hspace*{0.15in}
\includegraphics[scale=0.32]{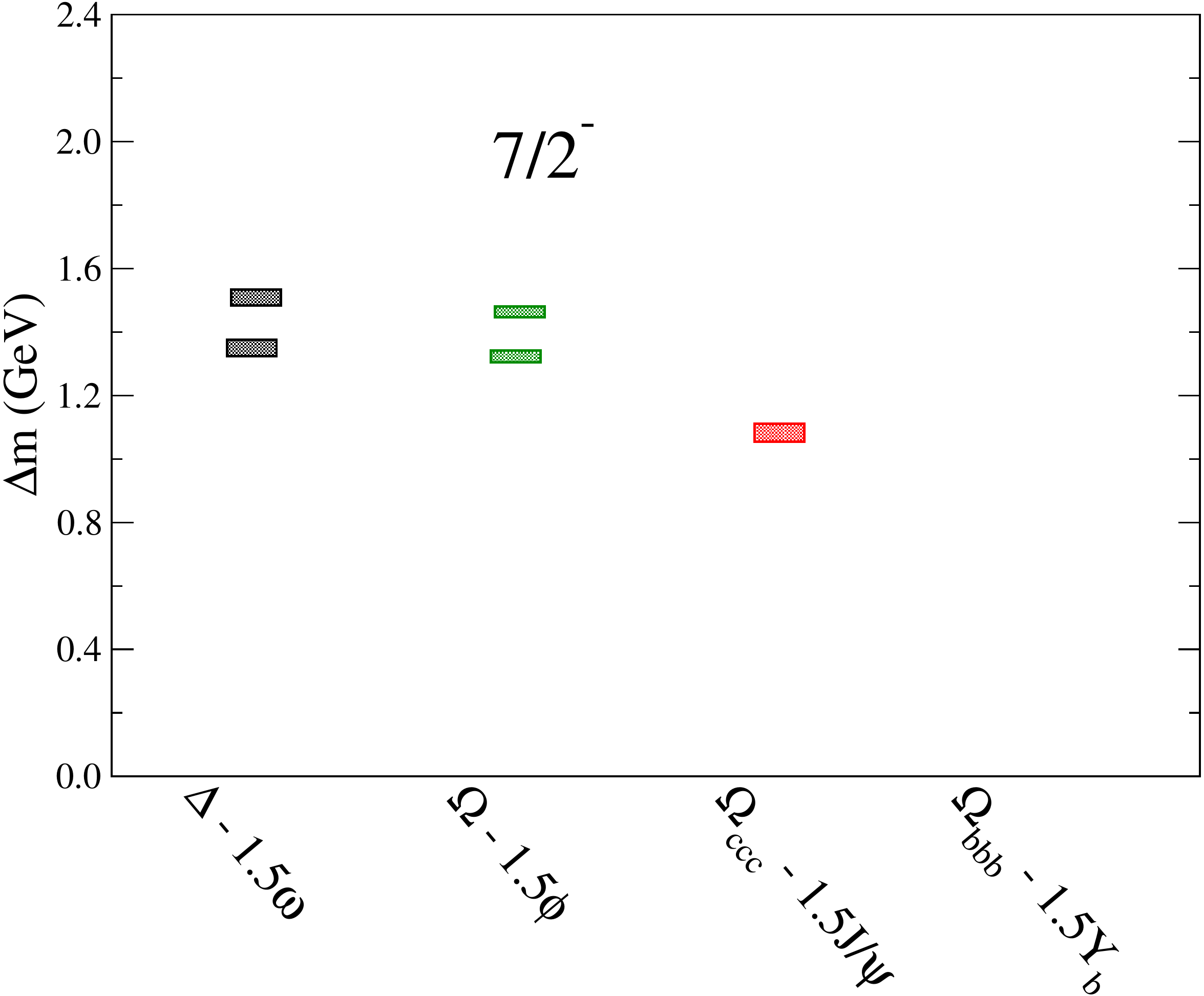}
\caption{Energy splittings of triply-flavoured baryons at various quark masses ($u$ to $b\, \rightarrow$  left to right) from the respective isoscalar vector meson (irrep $T_1^{--}$) ground state. Top four figures are for positive parity states and bottom four are for negative parity states respectively. A factor 3/2 is multiplied with isoscalar vector meson masses to account for the difference in the number of charm quarks  in baryons and mesons. For charm quark, results are from this work, while for light and strange quark results are from Ref.~\cite{Edwards:2012fx} and those for bottom quark are from Ref.~\cite{Meinel:2012qz}. The zebra-shaded boxes are the results obtained by using only non-relativistic operators.}
\hspace*{0.1in}
\eef{jpsipcomparison}

As presented in \tbn{n_operators}, operators with increasing numbers of
gauge-covariant derivatives (which can correspond to non-zero orbit angular 
momentum in a quark model) create states with numerous values of total spin 
$J$. For example, we construct flavour ($F$) decuplet states with
$D = 2, S = \frac32$ and $L = 2$ with the combination $\{10F_{S}
\otimes (S)_S \otimes (D)_{S}\}$, where $S$ in the subscript stands
for symmetric combinations, as defined in
Ref.~\cite{Edwards:2011jj}. In this way of construction we get four
quantum numbers with $J^{P} = \frac12^{+}, \frac32^{+}, \frac52^{+}$
and $\frac72^{+}$. In the heavy quark limit the spin-orbit interaction 
vanishes since the interaction term is inversely 
proportional to the square of the heavy quark mass. States corresponding to 
quantum numbers
($|L,S,J^{P}\rangle \, \equiv |2, \frac32, \frac12^{+}\rangle, |2,
\frac32, \frac32^{+}\rangle, |2, \frac32, \frac52^{+}\rangle$ and $|2,
\frac32, \frac72^{+}\rangle$) will thus be degenerate in the heavy quark
limit. Similarly, two states with quantum numbers $J^{P} =
\frac12^{-}$ and $\frac32^{-}$ with $L = 1$ and $S = \frac12$ will
also be degenerate in the heavy quark limit. In
\fgn{split_ccu_n_comp} we plot absolute values of energy differences
between energy levels which originate from the spin-orbit interaction
of the following ($L\,,S$) pairs : (2,$\frac32$ --in the left column),
(2,$\frac12$ --in the middle column) and (1,$\frac12$ --in the right column).
We plot these spin-orbit energy splittings at
various quark masses corresponding to following triple-flavoured
baryons : $\Delta_{uuu}$, $\Omega_{sss}$, $\Omega_{ccc}$ and
$\Omega_{bbb}$. We identified the states with these ($L,S$) pairs by finding the
 operators which incorporate these pairs and which have major overlaps to these 
states. These energy differences are obtained from the ratio of
jackknifed correlators, which in general, reduce errorbars by cancelling
correlation of these correlators. For bottom baryons we use data from
Ref.~\cite{Meinel:2012qz} and for light and strange baryon results
from Ref.~\cite{Edwards:2012fx} are used.

It should be noticed that energy levels shown for bottom baryons  
are obtained from only non-relativistic operators~\cite{Meinel:2012qz}. For 
charm baryons we show results from the full set of operators (including 
relativistic operators) as well as from only non-relativistic operators, and 
for light and strange baryons the splittings results are obtained also with the full set operators. 
In some cases we find that the inclusion of the relativistic operators increase
errorbars.  As one can observe that the degeneracy
between these states is more or less satisfied both for bottom 
and charm quarks. However, data with higher statistics 
is necessary to identify the
breaking of this degeneracy at charm quark. We will address this issue
in future. For $\Delta(uuu)$ and $\Omega(sss)$ some of these
splittings are non-zero. This is expected because of the presence of
the light quark masses in the denominator of the spin-orbit interaction
which enhance these splittings.

We also compared how energy splittings change between light and heavy baryons.  
Some of these, such as the hyperfine splitting, vanish in the heavy quark limit 
while others become constant.  However, most splittings tend to be 
higher at lighter quark masses where relativistic effects are prominent. 
We determined 
the energy difference between the triply-flavoured baryons with respect to
the isoscalar vector mesons with two constituents of the same flavour. To
make a comparison which is independent of the quark mass in the heavy quark
limit, we subtract $\frac32$ of the vector meson mass, where this factor 
simply takes account of the difference in the number of charm quarks between 
baryons and mesons.
Specifically we consider following splittings: $m_{\Delta_{uuu}} -
\frac32\,m_{\omega_{\bar{u}u}}\,, m_{\Omega_{sss}} - \frac32\,m_{\phi_{\bar{s}s}
}\,,
m_{\Omega_{ccc}} - \frac32\,m_{J/\psi_{\bar{c}c}}$ and $m_{\Omega_{bbb}} -
\frac32\,m_{\Upsilon_{\bar{b}b}}$.  
\fgn{jpsipcomparison} shows these splittings for positive and negative parity 
states at varying quark masses from the light $u,d$ quarks up to the $b$-quark
mass. For $\Delta^{++}(uuu)$ and $\Omega_{sss}$ baryons, results are from 
Ref.~\cite{Edwards:2012fx}, while for $\Omega_{bbb}$, we use results from 
Ref.~\cite{Meinel:2012qz}. 
The zebra-shaded boxes are the results obtained using only non-relativistic 
operators. These splittings mimic the binding energies of triple-flavoured 
states and thus it is interesting to compare these as a function of quark 
masses. In \fgn{jpsipcomparison_mpi2} we plot these splittings, for the 
ground-state and a few excitations as a function of the square of the 
pseudoscalar meson masses. Notice that most of the splittings in various spin 
parity channels decrease with quark mass. In the heavy quark limit, naively 
one can expand the mass of a heavy hadron, with $n$ heavy quarks,  as 
$M_{H_{nq}} = n~m_Q + A + B/m_Q + \mathcal{O}(1/m_{Q^2})$~\cite{Jenkins:1996de}. 
We expect that these energy splittings can also be expressed in the 
form $a + b/m_Q$ and similarly in the heavy quark limit with $a + b/m_{ps}$. 
Note this form is not expected to be valid for light hadrons. Using this
function we fitted the data obtained for 
$m_{\Omega_{bbb}} - \frac32\,m_{\Upsilon_{\bar{b}b}}\,, m_{\Omega_{ccc}} -
\frac32\,m_{J/\psi_{\bar{c}c}}$ and $m_{\Omega_{sss}} - \frac32\,m_{\phi_{\bar{s
}s}}$. Because of the very different behaviour in the chiral limit, 
the light quark point, $m_{\Delta_{uuu}} - \frac32\,m_{\omega_{\bar{u}u}}$ 
is excluded from the fit. 
For the cases where data for the bottom quark are not available (mainly for
negative parity cases), fitting is done using only the two data at 
charm and strange masses. 
While there is no good reason for the heavy-quark inspired functional form to 
model the data at the strange quark mass, a good 
fit is still found. We also extrapolate fit results to lighter pion masses and 
observe that for many cases they pass through the light quark points
though those points are {\it not} included in these fits. The fitted results
for the parameters $a$ and $b$ are tabulated in \tbn{ab_table}.
However, as mentioned before, one should not use this extrapolation to get 
splittings at lighter quark masses as is evident in many cases where data 
points fall significantly below the fit. 
It is worth noting that that the energy splittings for the spin-$\frac32^{+}$ 
and spin-$\frac12^{-}$ ground states are almost constant even when varying 
the quark mass from light to bottom. In fact, we observe better fit with a 
constant term for these splittings.
\bef[t!]
\centering
\includegraphics[scale=0.55]{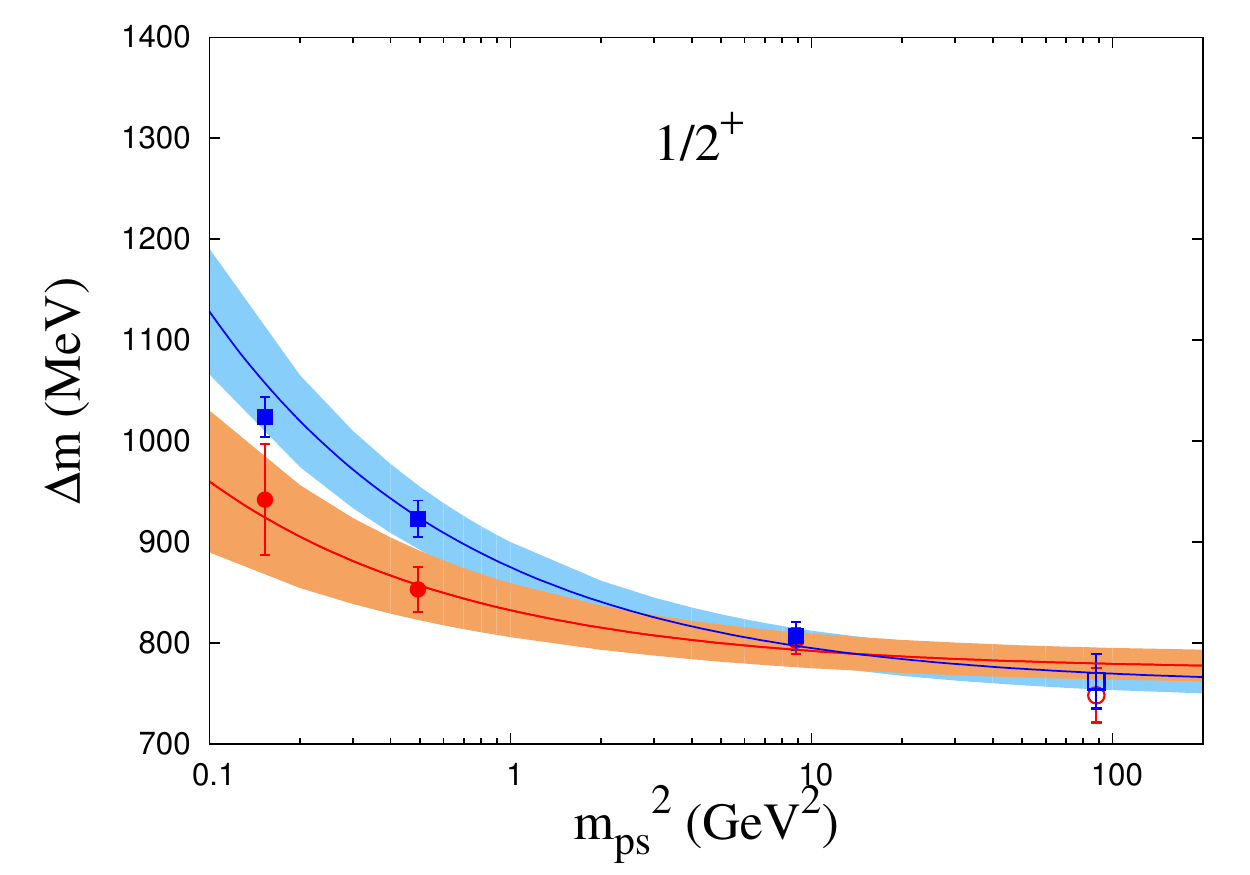}
\hspace*{0.1in}
\includegraphics[scale=0.55]{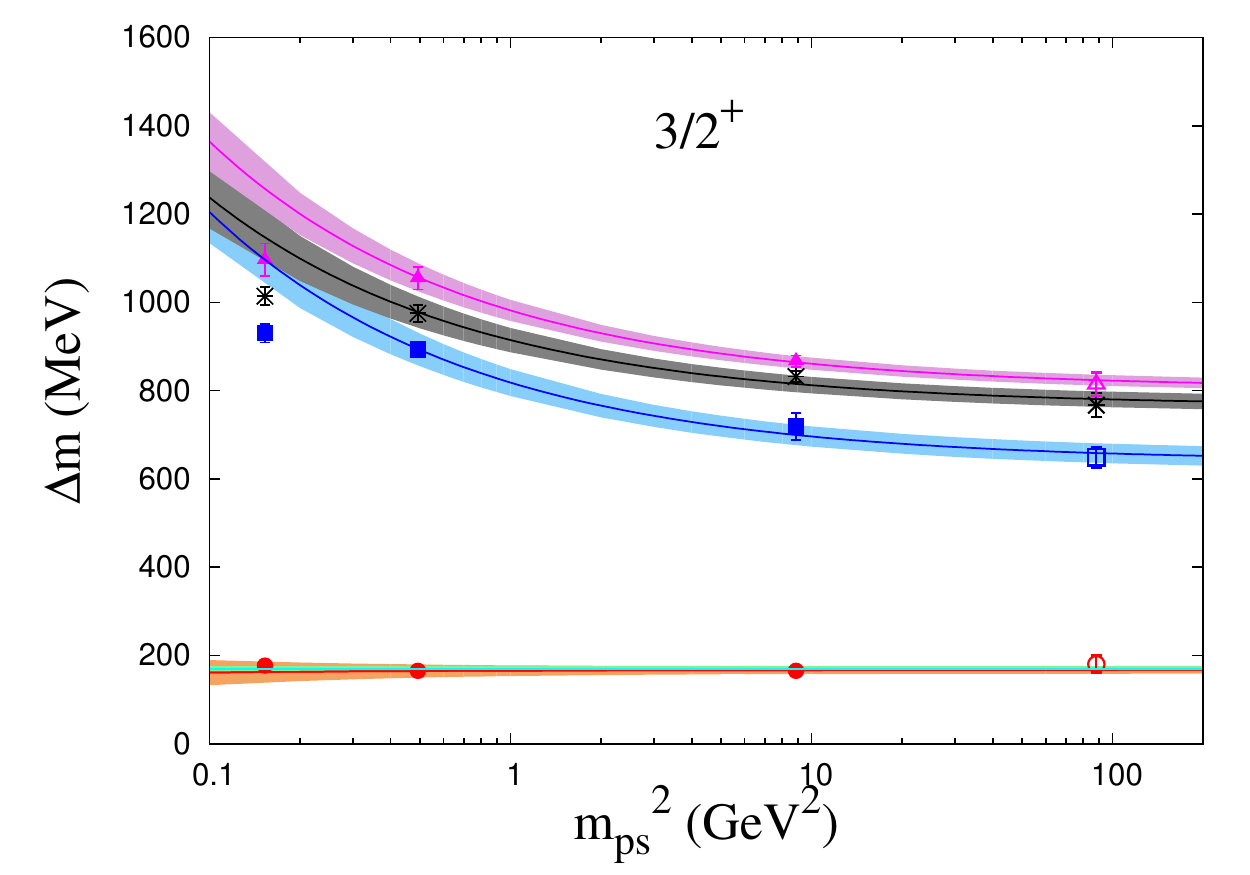}\\
\includegraphics[scale=0.55]{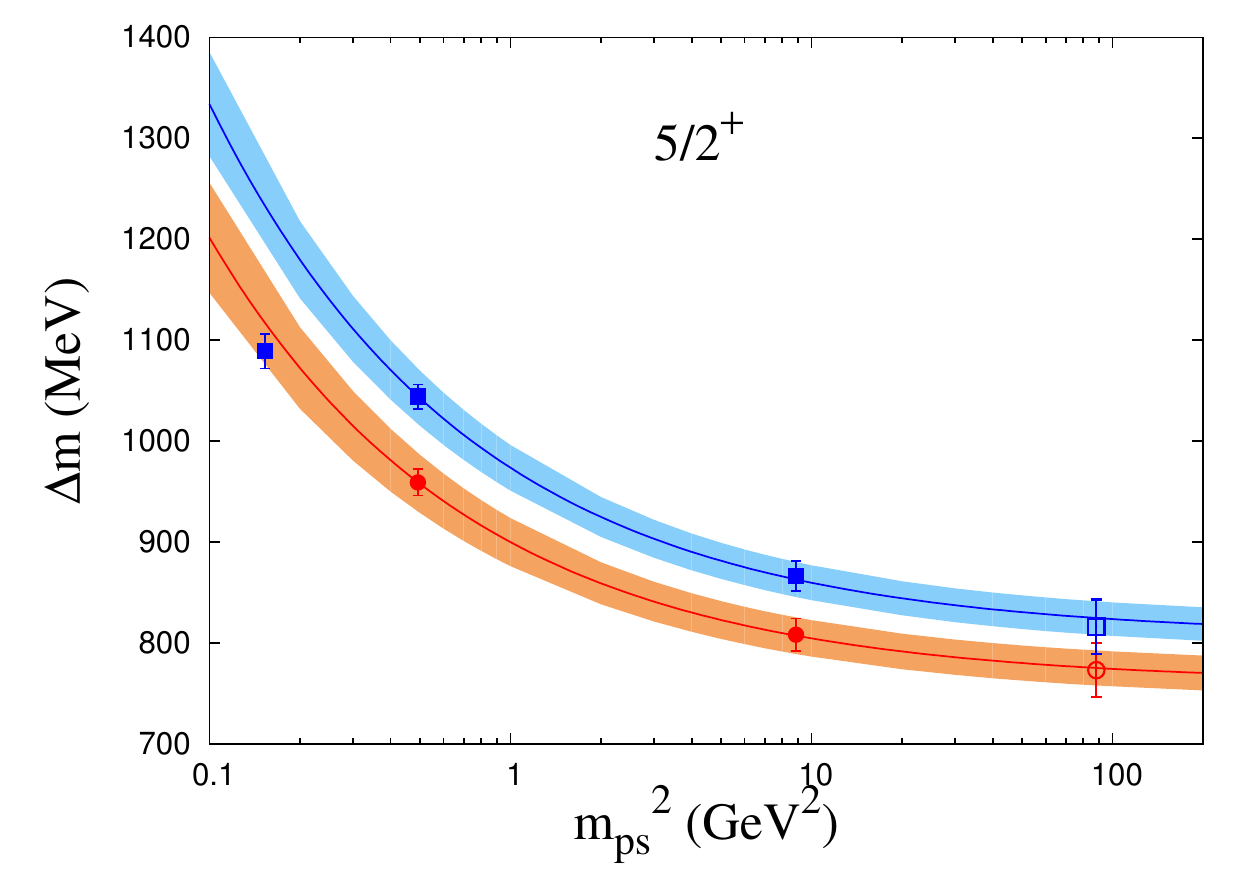}
\hspace*{0.1in}
\includegraphics[scale=0.55]{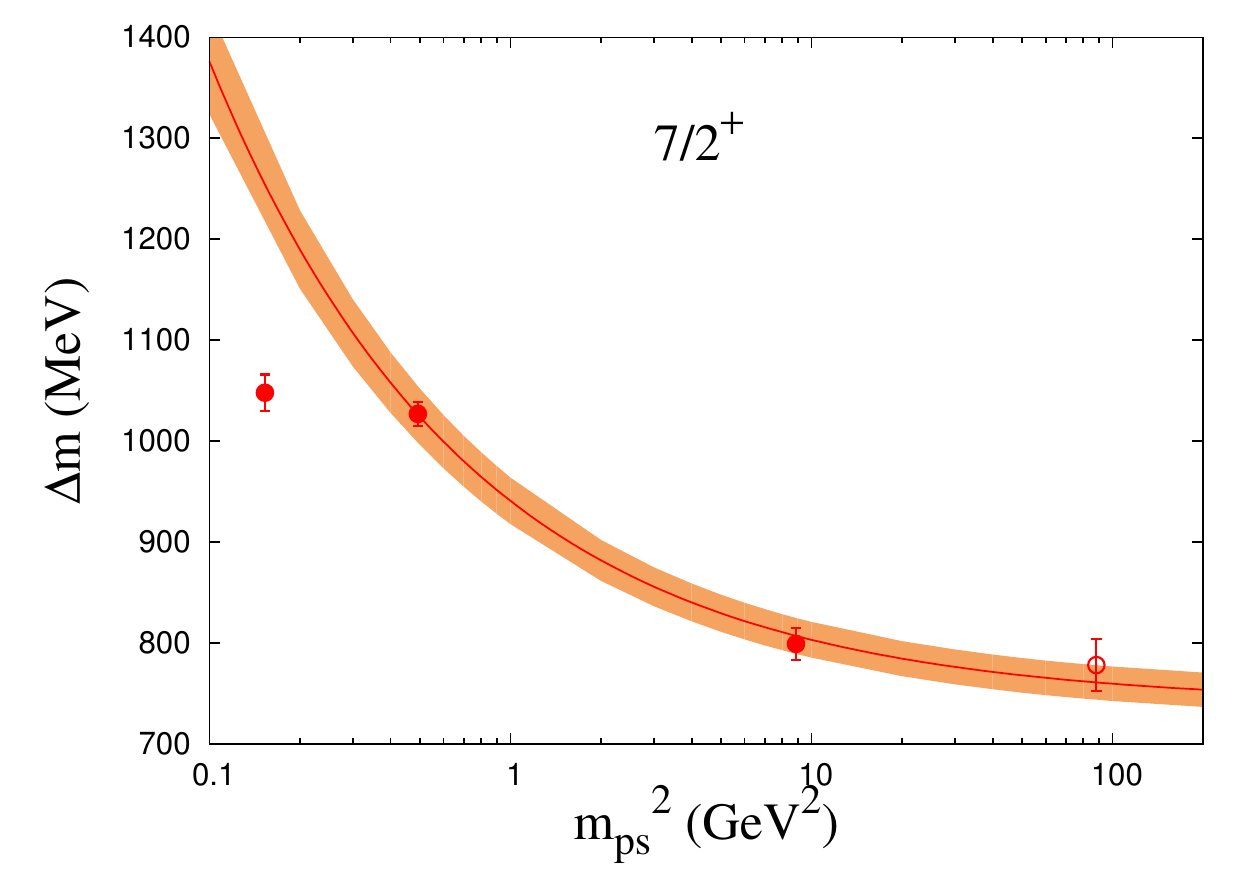}\\
\includegraphics[scale=0.55]{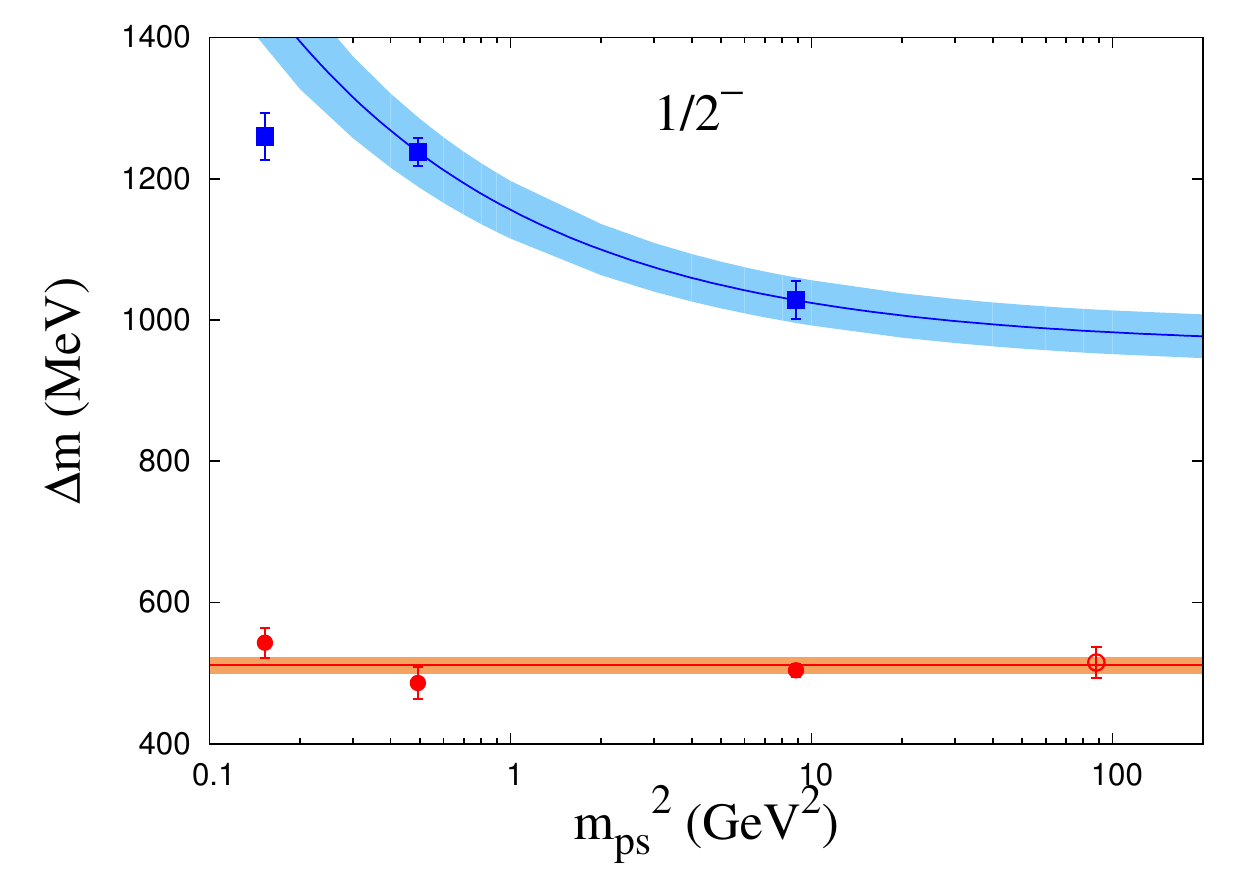}
\hspace*{0.1in}
\includegraphics[scale=0.55]{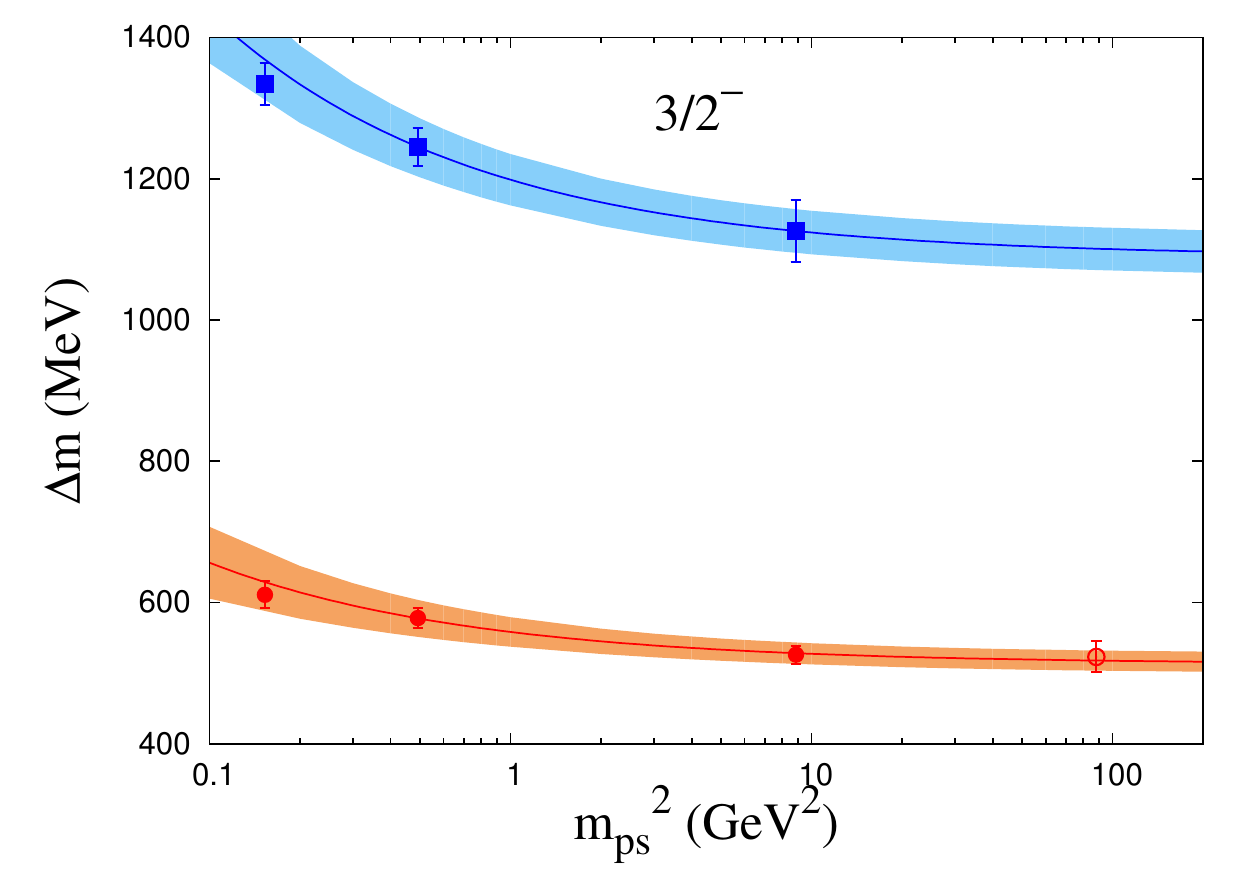}\\
\includegraphics[scale=0.55]{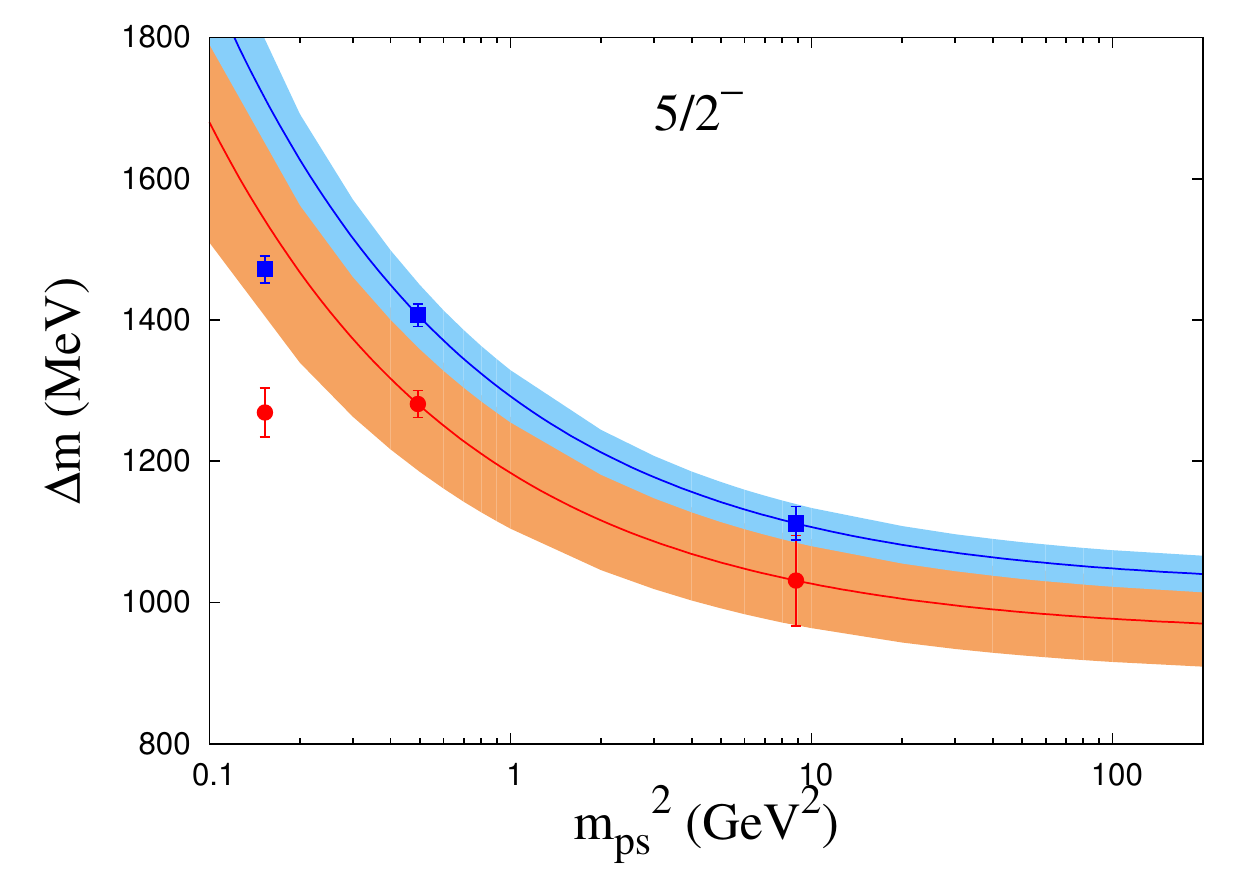}
\hspace*{0.1in}
\includegraphics[scale=0.55]{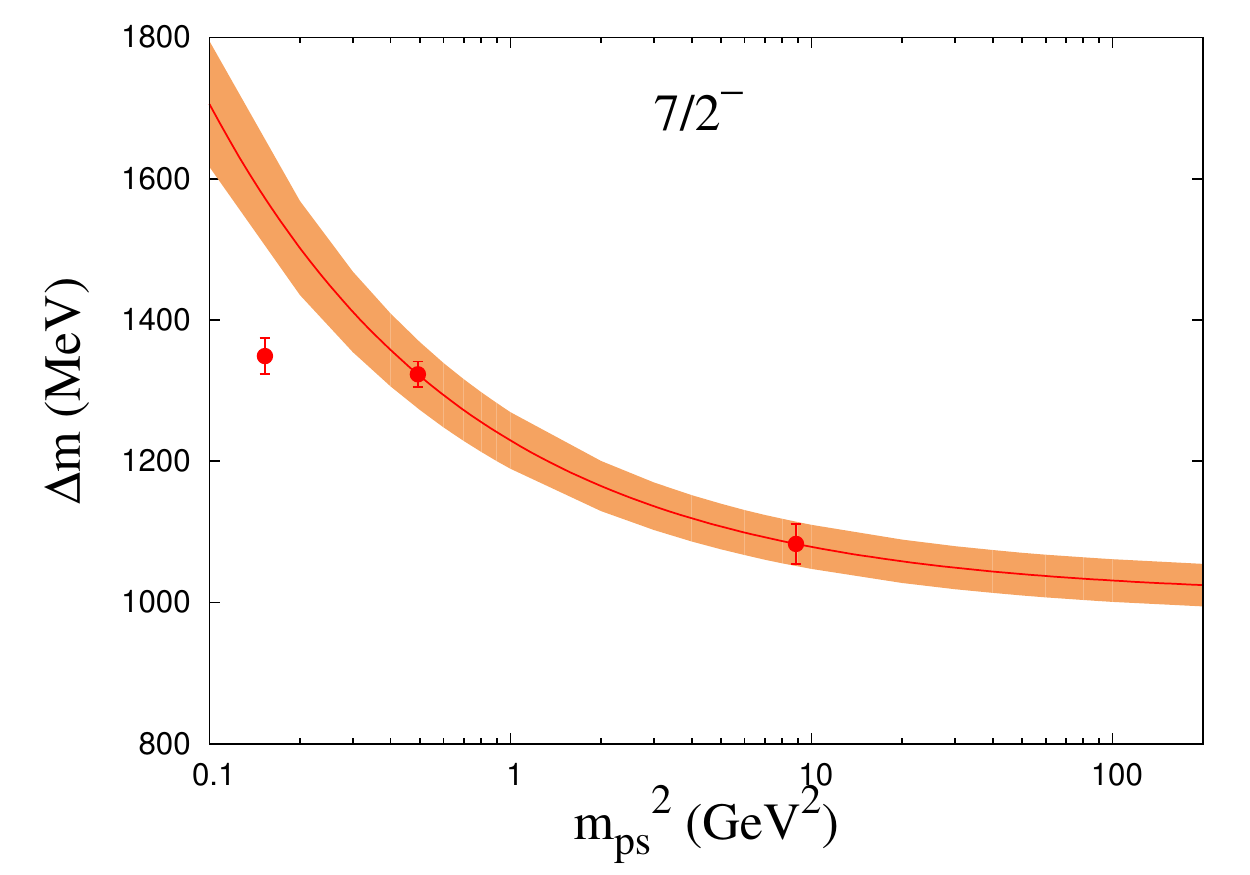}
\caption{Energy splittings of a few triple-flavoured baryons (labeled
  by different symbols and colours) from isoscalar vector meson (irrep
  $T_1^{--}$) ground state are plotted against the square of the
  pseudoscalar masses. 
Top four figures are for positive parity and
  bottom four are for negative parity baryons respectively. 
  This energy splitting mimics
  the binding energy and considering that these plots show the
  dependence of binding energies of these states as a function of
  quark mass. We fit this dependence with a form $a + b/m_{ps}$ (please see text for details).
}
\hspace*{0.1in}
\eef{jpsipcomparison_mpi2}
\bet[h]
\centering
\betb{ c | c | c | c  |||| c | c | c | c  c  c }
\hline 
\hline
State                              &    a      &     b      & $\cdof$ & State                              &    a      &     b      & $\cdof$ \\ \hline
$\mathbf{{\color{BurntOrange} \frac12^+}}$  & 773  (16) &  59   (22) & 1.8   & $\mathbf{{\color{BurntOrange} \frac12^-}}$  & 511  (12) &            & 0.9   \\
$\mathbf{{\color{SkyBlue} \frac12^+}}$      & 758  (16) & 117   (19) & 0.4   & $\mathbf{{\color{SkyBlue} \frac12^-}}$      & 963  (31) & 193   (27) &    \\ \hline
$\mathbf{{\color{BurntOrange} \frac32^+}}$  & 168   (9) &  -2.0  (9) & 0.5   & $\mathbf{{\color{BurntOrange} \frac32^-}}$  & 513  (14) &  45   (15) & 0.08   \\
$\mathbf{{\color{GreenYellow} \frac32^+}}$  & 170   (5) &            & 0.4   &                                             &           &            &         \\
$\mathbf{{\color{SkyBlue} \frac32^+}}$      & 640  (22) & 179   (21) & 0.56   & $\mathbf{{\color{SkyBlue} \frac32^-}}$      & 1089 (30) & 109   (20) &    \\
$\mathbf{{\color{Gray} \frac32^+}}$         & 765  (17) & 149   (21) & 1.2   &                                             &           &            &         \\
$\mathbf{{\color{CarnationPink} \frac32^+}}$& 805  (12) & 177   (21) & 0.18   &                                             &           &            &         \\ \hline
$\mathbf{{\color{BurntOrange} \frac52^+}}$  & 760  (17) & 140   (16) & 0.01   & $\mathbf{{\color{BurntOrange} \frac52^-}}$  & 954  (61) & 230   (50) &    \\
$\mathbf{{\color{SkyBlue} \frac52^+}}$      & 807  (17) & 167   (15) & 0.15   & $\mathbf{{\color{SkyBlue} \frac52^-}}$      & 1021 (26) & 271   (26) &    \\ \hline
$\mathbf{{\color{BurntOrange} \frac72^+}}$  & 739  (17) & 201   (16) & 0.7   & $\mathbf{{\color{BurntOrange} \frac72^-}}$  & 1009 (30) & 220   (27) &    \\ \hline
\hline
\eetb
\caption{Fitted values of the parameters $a$ and $b$ for the heavy quark expansion of the mass difference $\Delta m_H = a + b/m_{ps}$. Left side is for positive parity and the right side is for negative parity states. Colour coding for these states are corresponding to \fgn{jpsipcomparison_mpi2}. We do not quote $\cdof$ for the cases where there are only two data points ({\it i.e.}, data points are not available at the bottom quark).}
\eet{ab_table}

\section{Conclusions}
In this work, we present results from the first non-perturbative
calculation on the excited state spectroscopy of the triply-charmed
baryons with spin up to 7/2.  This system is the baryonic analogue of
charmonium, which has been central to understanding more about the
phenomenology of the strong force.  Calculations were performed using
anisotropic lattice QCD with a background of $2+1$ dynamical light
quarks with pion mass of about 390 MeV in a finite volume with extent
of approximately 1.9 fm where the temporal and spatial spacings were
$a_{t} = 0.0351(2)$ fm and $a_s \sim 0.12$ fm respectively. The systematic uncertainties due to chiral, continuum and final volume extrapolations have not been explored here and those will be addressed in future.

Using the distillation smearing technique to make use of the variational 
method, we extracted the 
spectrum of triply-charmed baryons comprising a total of 18 states, including 
spin up to 7/2.  We use a large set of continuum operators which can be 
classified according to the irreducible representation of $SU(3)_F$ flavour 
and which transform according to good total angular momentum in the continuum. 
Angular momenta are realized by including up to two covariant derivatives in 
creation operators.  
Similar to earlier work, we observe approximate rotational 
symmetry for these operators at the scale of hadrons. By using this symmetry 
and calculating 
overlap factors of various operators to energy eigenstates and then comparing
those over irreps we are able to extract states reliably with spin up
to 7/2.  We also 
observe that 
there is strong mixing between states created by non-relativistic non-hybrid 
and relativistic non-hybrid type operators but comparatively weaker mixing 
between those created by hybrid and non-hybrid type operators. 
Additional suppression of mixing is 
seen for non-relativistic operators with given $J$ but with different $L$
and $S$ compared to those with the same $J, L$ and
$S$. However, this suppression is not present for relativistic
operators.

The main results are shown in \fgn{3o2etc_splitting_spectrum_mev}. 
 As in Ref.~\cite{Edwards:2012fx}, we also
find bands of states with alternating parities and increasing
energies. Beside identifying the spin of a state we are also able to
decode the structure of operators leading to that state : whether
constructed by relativistic, non-relativistic, hybrids, non-hybrid
types or a mixture of them all. However, for negative parity states
and highly excited positive parity states, this identification is not
possible since we do not include operators with more than two
derivatives which will contribute to these states and so will change
the relative contribution from various operators leading to such
states. Similar to light and strange baryon
spectra~\cite{Edwards:2012fx}, we also find the number of extracted
states of each spin in the three lowest-energy bands and the number of
quantum numbers expected based on weakly broken $SU(6)\times O(3)$
symmetry agree perfectly, {\it i.e.,} the triply-charmed baryon
spectra remarkably resemble the expectations of quantum numbers from
quark model~\cite{Greenberg:1964pe,Isgur:1977ef,Isgur:1978xj}. Even
the inclusion of non-relativistic hybrid operators does not spoil this
agreement for positive parity states. However, with the inclusion of
relativistic operators which contribute more to the higher excited
states it is expected that this band structure will not be
followed. That is what we also observe for the higher excited states.
However, it is to be noted that we have not used any multi-hadron
operators in this calculation. Inclusion of those operators, particularly those involving light quarks, may affect some of the above conclusions, though to a lesser extent than their influence in the light hadron spectra.

Various energy splittings, including splittings due to spin-orbit
coupling, are also evaluated for these baryons and those are compared with similar splittings obtained at light, strange as
well as bottom quark masses. While the data at charm quark mass (for
$\Omega_{ccc}$) is from this work, those for light and strange quark masses
(for $\Delta_{uuu}$ and $\Omega_{sss}$ baryons) are taken from
Ref.~\cite{Edwards:2012fx} and bottom quark results (for
$\Omega_{bbb}$) are from Ref.~\cite{Meinel:2012qz}. The energy
splittings between baryons due to spin-orbit coupling vanish at the
heavy quark limit.  To check this degeneracy we identify the states
with same $L$ and $S$ values from overlap factors of various operators
and find that the spin-orbit energy degeneracy between these states
are more or less satisfied both for bottom and charm quarks. However,
for a few cases they are non-zero at lighter quark masses.  More
precise data is necessary to check the breaking of this
degeneracy at the charm quark mass.

The energy splitting of the triply-charmed baryon spectrum,
from the isoscalar vector meson (irrep $T_1^{--}$) ground state 
is also evaluated. These splittings are compared with similar ones 
obtained at other quark masses. For the splitting, which mimics the binding 
energy 
of these states, significant quark mass dependence is observed for ground as 
well as for first few excited states, except for the 
$J^{P} = 3/2^{+}$ and $J^{P} = 1/2^{-}$ ground states. These splittings 
can be modelled with a form $a + b/m_{\rm ps}$ to show their expected quark 
mass dependence which assumes they will tend to a constant in the heavy quark 
limit, 
It is interesting to note this form gives a good fit with data at bottom, 
charm as well as strange masses. For some of them, we observe that the 
extrapolated fit lines pass through the light quark data points even when 
they are not included in the fit.

\section{Acknowledgements}
We thank our colleagues within the Hadron Spectrum Collaboration.  NM
also likes to thank S. Datta for useful
discussions. Chroma~\cite{Edwards:2004sx} and
QUDA~\cite{Clark:2009wm,Babich:2010mu} were used to perform this work
on the Gaggle and Brood clusters of the Department of Theoretical
Physics, Tata Institute of Fundamental Research and at Lonsdale cluster
maintained by the Trinity Centre for High Performance Computing funded
through grants from Science Foundation Ireland (SFI), at the SFI/HEA
Irish Centre for High-End Computing (ICHEC) and at Jefferson
Laboratory under the USQCD Initiative and the LQCD ARRA project.
Gauge configurations were generated using resources awarded from the
U.S. Department of Energy INCITE program at the Oak Ridge Leadership
Computing Facility at Oak Ridge National Laboratory, the NSF Teragrid
at the Texas Advanced Computer Center and the Pittsburgh Supercomputer
Center, as well as at Jefferson Lab.  MP acknowledges support from the Trinity College Dublin Indian Research Collaboration Initiative 
and the Council of Scientific and Industrial Research, India for financial support through 
the Shyama Prasad Mukherjee Fellowship; 
NM acknowledges support from Department of Science and Technology, India under 
grant No. DST-SR/S2/RJN-19/2007; RGE acknowledges support from support from 
U.S. Department of Energy contract DE-AC05-06OR23177, under which Jefferson 
Science Associates, LLC manages and operates Jefferson Laboratory. 
This research was supported by the European Union under Grant Agreement number 
238353 (ITN STRONGnet).



\end{document}